\documentclass[12pt,aps,prd,preprint,
tightenlines,nosuperscriptaddress,amsmath,amssymb,nofootinbib]{revtex4-1} 

\RequirePackage[colorlinks=true
,urlcolor=blue
,anchorcolor=blue
,citecolor=blue
,filecolor=blue
,linkcolor=blue
,menucolor=blue
,pagecolor=blue
,linktocpage=true
,pdfproducer=medialab
,pdfa=true
]{hyperref}

\allowdisplaybreaks

\usepackage{amsmath,amssymb,amsthm,amsfonts}
\usepackage{graphicx}
\usepackage{subfigure}
\usepackage{dcolumn}	
\usepackage{hyperref}      	
\usepackage{bm}		
\usepackage{epsfig}
\usepackage{epstopdf}
\usepackage{setspace}
\usepackage[usenames, dvipsnames]{color}
\usepackage{slashed}
\usepackage{comment}
\usepackage{enumitem}
\usepackage{siunitx}
\usepackage{multirow}
\usepackage{feynmp}
\usepackage{datetime}
\usepackage{enumitem}
\usepackage{titlesec, titletoc}
\titleformat*{\section}{\boldmath\bfseries}
\titleformat*{\subsection}{\boldmath\bfseries}
\setlist[description]{leftmargin=0.4cm}

\makeatletter
\def\endfmffile{%
	\fmfcmd{\p@rcent\space the end.^^J%
		end.^^J%
		endinput;}%
	\if@fmfio
	\immediate\closeout\@outfmf
	\fi
	\ifnum\pdfshellescape>\z@
	\immediate\write18{mpost \thefmffile}%
	\fi}
\makeatother

\newcommand{\PRE}[1]{{#1}} 

\newcommand{\be}{\begin{equation}\begin{aligned}}
\newcommand{\ee}{\end{aligned}\end{equation}}
\newcommand{\beq}{\begin{equation}}
\newcommand{\eeq}{\end{equation}}
\newcommand{\beqa}{\begin{eqnarray}}
\newcommand{\eeqa}{\end{eqnarray}}

\newcommand{\ev}{\text{eV}}

\newcommand{\mev}{\text{MeV}}

\renewcommand{\eqref}[1]{Eq.~(\ref{eq:#1})}
\newcommand{\eqsref}[2]{Eqs.~(\ref{eq:#1}) and (\ref{eq:#2})}
\newcommand{\secref}[1]{Sec.~\ref{sec:#1}}

\newcommand{\figref}[1]{Fig.~\ref{fig:#1}}
\newcommand{\figsref}[2]{Figs.~\ref{fig:#1} and \ref{fig:#2}}

\newcommand{\tableref}[1]{Table~\ref{table:#1}}

\RequirePackage[normalem]{ulem}

\begin{document}
 
\count\footins = 1000
 
\preprint{UCI-TR-2020-01}

\title{
\PRE{\vspace*{1in} }
{\Large Dynamical Evidence For a Fifth Force Explanation \\
of the ATOMKI Nuclear Anomalies
} 
\PRE{\vspace*{0.5in} }
}

\author{Jonathan~L.~Feng}
\email{jlf@uci.edu}
\affiliation{Department of Physics and Astronomy, 
University of California, Irvine, CA 92697-4575, USA
\PRE{\vspace*{0.5in}}
}

\author{Tim~M.~P.~Tait}
\email{ttait@uci.edu}
\affiliation{Department of Physics and Astronomy, 
University of California, Irvine, CA 92697-4575, USA
\PRE{\vspace*{0.5in}}
}

\author{Christopher~B.~Verhaaren\PRE{\vspace*{0.1in}}
}
\email{cverhaar@uci.edu}
\affiliation{Department of Physics and Astronomy, 
University of California, Irvine, CA 92697-4575, USA
\PRE{\vspace*{0.5in}}
}

\begin{abstract}
\PRE{\vspace*{0.2in}}
Recent anomalies in $^8$Be and $^4$He nuclear decays can be explained by postulating a fifth force mediated by a new boson $X$.  The distributions of both transitions are consistent with the same $X$ mass, 17 MeV, providing kinematic evidence for a single new particle explanation.  In this work, we examine whether the new results also provide dynamical evidence for a new particle explanation, that is, whether the observed decay rates of both anomalies can be described by a single hypothesis for the $X$ boson's interactions. We consider the observed $^8$Be and $^4$He excited nuclei, as well as a $^{12}$C excited nucleus; together these span the possible $J^P$ quantum numbers up to spin 1.   For each transition, we determine whether scalar, pseudoscalar, vector, or axial vector $X$ particles can mediate the decay, and  we construct the leading operators in a nuclear physics effective field theory that describes them.  Assuming parity conservation, the scalar case is excluded and the pseudoscalar case is highly disfavored.  Remarkably, however, the protophobic vector gauge boson, first proposed to explain only the $^8$Be anomaly, also explains the $^4$He anomaly within experimental uncertainties.  We predict signal rates for other closely related nuclear measurements, which, if confirmed, will provide overwhelming evidence that a fifth force has been discovered.  
\end{abstract}


\maketitle
\tableofcontents


\clearpage

\section{Introduction}
\label{sec:introduction}

In the standard model of particle physics, particles interact through the electromagnetic, strong, and weak forces.  Following tradition, if we include gravity, but not the force mediated by the Higgs boson, there are four known forces.  The discovery of a fifth force and the particle that mediates it would have profound consequences for fundamental physics. 

Nuclear physics provides a fruitful hunting ground for new forces. This was obviously true in the past, given the central role of nuclear physics in elucidating the strong and weak forces, but it remains true today. In particular, nuclear experiments are well suited to discovering light and weakly interacting particles~\cite{Treiman:1978ge,Donnelly:1978ty,Freedman:1984sd,Savage:1986ty}, which have many particle and astrophysical motivations~\cite{Holdom:1985ag,Boehm:2003hm,Pospelov:2007mp,Feng:2008ya}.  These particles may be light enough to be produced in nuclear decays, and the extraordinary event rates of nuclear experiments allow them to probe extremely rare processes mediated by particles with very weak interactions.

In 2015, researchers working at the ATOMKI pair spectrometer experiment reported a 6.8$\sigma$ anomaly in the decays of an excited beryllium nucleus to its ground state, $^8\text{Be(18.15)} \to {}^8\text{Be} \, e^+ e^-$~\cite{Krasznahorkay:2015iga}.  The observed anomaly is an excess of events at opening angles $\theta_{e^+e^-} \approx 140^{\circ}$, suggesting the production of a new $X$ boson through $^8\text{Be(18.15)} \to {}^8\text{Be} \, X$, followed by the decay $X \to e^+ e^-$, with best-fit parameters $m_X = 16.7 \pm 0.35 \, (\text{stat}) \pm 0.5 \, (\text{sys})~\mev$ and $\Gamma (^8\text{Be(18.15)} \to {}^8\text{Be} \, X) = 1.1 \times 10^{-5}~\text{eV}$, assuming $B(X \to e^+ e^-) = 1$.  

In 2016, this possibility was analyzed in detail in Refs.~\cite{Feng:2016jff,Feng:2016ysn}.  All possible spin-parity assignments for the $X$ particle were examined, several explanations were excluded, and the possibility that the $X$ particle is a vector gauge boson mediating a protophobic fifth force was shown to be a viable explanation.  Following these studies, other new physics explanations were examined in detail, with a host of implications for nuclear, particle, and astrophysical observations; see, for example, Refs.~\cite{Gu:2016ege,Liang:2016ffe,Kitahara:2016zyb,Ellwanger:2016wfe,Kahn:2016vjr,Seto:2016pks,Fayet:2016nyc,Kozaczuk:2016nma,DelleRose:2017xil,Alves:2017avw,DelleRose:2018eic,Chen:2019ivz,Nam:2019osu,Hati:2020fzp} and Ref.~\cite{Fornal:2017msy} for a brief review.  A follow up study from nuclear theory was also shown to disfavor a nuclear form factor as an explanation~\cite{Zhang:2017zap}.  Follow up measurements of the nuclear transitions have not been reported by other groups, but other experimental collaborations have provided new constraints on $X$ boson couplings to electrons that exclude part of the viable parameter space in some models~\cite{Banerjee:2018vgk,Banerjee:2019hmi}. 

Since 2016, the ATOMKI experimentalists have re-built their spectrometer and confirmed the $^8\text{Be(18.15)}$ anomaly with the new detector.  A refined analysis yields best-fit parameters~\cite{Krasznahorkay:2018snd} 
\begin{eqnarray}
m_X  &=&  17.01 \pm 0.16\, \mev \ , \\
\Gamma (^8\text{Be(18.15)} \to {}^8\text{Be} \, X ) 
 &=&  (6 \pm 1)   \times  10^{-6} \, \Gamma (^8\text{Be(18.15)}   \to  {}^8\text{Be} \, \gamma )  \nonumber \\
 &= & (1.2 \pm 0.2)  \times  10^{-5}\, \text{eV} \ .  \label{eq:Bewidth}
\end{eqnarray}
They have also considered the decays of other excited beryllium and carbon nuclei~\cite{Krasznahorkay:2017bwh,Krasznahorkay:zakopaneabs}.  Most remarkably, in recent months they have reported the observation of a 7.2$\sigma$ excess at $\theta_{e^+e^-} \approx 115^{\circ}$ in the transitions of $^4\text{He}(20.49) \to {}^4\text{He} \, e^+ e^-$, where the proton beam energy was tuned to produce a resonance that sits within the Breit-Wigner peaks of both the $^4$He(20.21) and $^4$He(21.01) excited states.  Assuming the production of a new particle, the best fit parameters are reported to be~\cite{Krasznahorkay:2019lyl,Krasznahorkay_He_Proceedings}
\begin{eqnarray}
m_X  &=&  16.98 \pm 0.16 \, (\text{stat}) \pm 0.20 \, (\text{sys}) \, \mev \ ,  \\
\Gamma (^4\text{He}(20.49) \to {}^4\text{He} \,  X)  &=&  0.12 \ \Gamma( ^4\text{He}(20.21) \to {}^4\text{He} \, e^+ e^-)_{\text{E0}} \nonumber \\
&=& (4.0 \pm 1.2) \times 10^{-5} \,\text{eV} \, , \label{eq:width}
\end{eqnarray}
where the width uncertainty includes only the uncertainty from $\Gamma( ^4\text{He}(20.21) \to {}^4\text{He} \, e^+ e^-)_{\text{E0}} =(3.3\pm1.0)\times 10^{-4}~\ev$~\cite{Walcher:1970vkv}, the decay width of the standard model E0 transition. The fact that this excess is at a different angle eliminates some experimental systematic explanations of both anomalies.  At the same time, the fact that the excess is found at the same $X$ mass provides striking supporting evidence for a new particle is being produced. 

Of course, it would be clarifying if these nuclear decays were examined by another collaboration. It is important to note, however, that the observation of a second signal provides an opportunity for another highly non-trivial check of the new physics interpretation beyond simply the consistency of the best-fit masses. The description in terms of a new particle requires not just kinematic consistency, but also dynamical consistency; that is, the rates for the anomalous decays must also be consistent with the same set of interaction strengths between the $X$ and quarks.  The best-fit widths are similar numerically, but it is far from obvious that this implies consistency: the $^8$Be and $^4$He excited states have different $J^P$ quantum numbers and different excitation energies, and the $^8$Be measurement was on resonance, while the $^4$He measurement was between two resonances.  The decays therefore take place through operators with different dimensions and different partial waves, and these differences may lead to disparate decay widths.  {\em A priori}, then, it is not at all clear that the observed anomalous $^8$Be and $^4$He decay rates have a consistent new physics interpretation.

On the experimental side, the width measurement of \eqref{width} requires careful differentiation of signal and background.  The leading backgrounds include external pair creation ($p+{}^3\text{H}\to {}^4\text{He} + (\gamma^\ast \to e^+ e^-)$) and internal pair creation from the E0 transition ${}^4\text{He}(20.21) \to {}^4\text{He} \, e^+e^-$.  By fitting to the expected opening angle distributions for the signal and these backgrounds, the ratio $\Gamma_X/\Gamma _{\text{E0}}$ was determined.  The collaboration has noted, however, that an additional background from the E1 transition may also be significant~\cite{KrasznahorkayCommunication}.  Including this contribution will modify the best-fit width. With this in mind, we consider models that predict widths that differ by orders of magnitude from \eqref{width} to be highly disfavored, whereas models that predict roughly similar widths are considered viable.  These predictions will be tested once a refined experimental analysis including a separate determination of the E1 background is available.

We begin examining the fifth force explanation for the ATOMKI results in \secref{kinematics}, where we consider the production and decay kinematics for the observed $^8$Be and $^4$He processes and also for a $^{12}$C excited state, which completes the possible nuclei $J^P$ assignments up to spin~1.  In \secref{spinparity} we determine how the observation or exclusion of various decays constrains the possible $X$ spin-parity assignments. This is developed further in \secref{OpAnalysis}, where for each of these possibilities we determine the leading interactions in an effective field theory (EFT) that describes low-energy nuclear transitions.  Using this EFT, the decay rates for the pseudoscalar, axial vector, and vector cases are calculated in Secs.~\ref{sec:pseudoscalar}, \ref{sec:axialvector}, and \ref{sec:VecX}, respectively.  Remarkably, we find that in the vector case, the protophobic gauge boson, previously advanced as an explanation of the $^8$Be anomaly, also provides a viable explanation for the new $^4$He observations within experimental uncertainties.  In \secref{general}, we suggest further nuclear measurements that can provide incisive tests of the new particle hypothesis, and we summarize our conclusions in \secref{conclusions}.

\section{Production and Decay Kinematics}
\label{sec:kinematics}

We consider the production of new $X$ particles in the decays of excited states of $^8$Be, $^{12}$C, and $^4$He nuclei.  
Although the states of interest have different quantum numbers, leading to different decay dynamics, the kinematic features of the relevant processes are very similar.  
In each case, a beam of protons with kinetic energy $E_{\text{beam}}$ collides with nuclei $A$ at rest to form excited nuclei through the 
process $p + A \to N_*$.  The excited nuclei $N_*$ decay to the corresponding ground state nuclei $N_0$ through $N_* \to N_0 X$, and the $X$ bosons further 
decay via $X \to e^+ e^-$.  In this section we review the experimental results and the kinematics of these processes.  
We examine the kinematics of resonance production in \secref{production}, and consider the $X$ boson decay and the 
electron-positron opening angle $\theta_{e^+ e^-}$ in \secref{decay}.

\subsection{Resonance Production}
\label{sec:production}

\begin{table}[tbp]
\caption{Production and decay kinematic parameters.  Beams of protons with kinetic energy $E_{\text{beam}}$ collide with nuclei $A$ at rest to form excited nuclei $N_*$, which then decay to the ground state nucleus $N_0$ through $N_* \to N_0 X$.  We fix $m_X = 17~\mev$ and for each of the relevant processes, we give the values of $E_{\text{beam}}$, $m_A$, $m_{N_*}$, $v_{N_*}$ (the $N_*$ velocity in the lab frame), $v_X$ (the $X$ velocity in the $N_*$ rest frame), and $\theta_{e^+ e^-}^{\text{min}}$ (the minimum $e^+e^-$ opening angle). $^4$He(20.49) indicates the resonance energy probed in Ref.~\cite{Krasznahorkay:2019lyl}, which sits between the $^4$He(21.01) and $^4$He(20.21) states. 
\vspace*{-.05in}
 \label{table:kinematics} }
\begin{tabular}[t]{rclccccccc}
 \toprule 
\rule[-0.2cm]{0pt}{0.6cm} $p + \ A\ $ & $\to$ & \ \quad $N_*$  & \ $E_{\text{beam}}$ (MeV) \ & \ $m_A$ (MeV) \ & \ $m_{N_*}$ (MeV) \ & \  $v_{N_*}/c$ \ & \ \ $v_X/c$ \ \ & \ $\ \theta_{e^+ e^-}^{\text{min}}$ \\
\hline
\rule{0pt}{0.4cm} $p + \text{$^7$Li}$ & $\to$ & $^8$Be(18.15) & 1.03 & 6533.83 &  7473.01  & 0.0059 & 0.350 & 139$^\circ$ \\
\rule{0pt}{0.4cm} $p + \text{$^7$Li}$ & $\to$ & $^8$Be(17.64) & 0.45 & 6533.83 &7472.50 & 0.0039 & 0.267 &149$^\circ$ \\
\hline
\rule{0pt}{0.4cm} $p + \text{$^{11}$B}$ & $\to$ & $^{12}$C(17.23) & 1.40 & 10252.54 & 11192.09 & 0.0046 & 0.163 & 161$^\circ$ \\
\hline
\rule{0pt}{0.4cm} $p + \text{$^3$H}$ & $\to$ & $^4$He(21.01) & 1.59 & 2808.92 &  3748.39 & 0.0146 & 0.587 & 108$^\circ$ \\
\rule{0pt}{0.4cm} $p + \text{$^3$H}$ & $\to$ & $^4$He(20.49) & 0.90 & 2808.92 &  3747.87 & 0.0110 & 0.557 & 112$^\circ$ \\
\rule{0pt}{0.4cm} $p + \text{$^3$H}$ & $\to$ & $^4$He(20.21) & 0.52 & 2808.92 &  3747.59 & 0.0084 & 0.540 & 115$^\circ$ \\
\toprule
 \end{tabular}
\vspace*{0.15in}

\caption{Nuclear excited states $N_*$ and their spin-parity $J_*^{P_*}$, isospin $T_*$, total decay width $\Gamma_{N_*}$, and photon branching fraction~\cite{nndc,Tilley:1992zz,Tilley:2004zz,Kelley:2017qgh,Wiringa:2013fia,Segel:1965zz} (or, in the case of $^4$He(20.21), the branching fraction for the E0 decay into $e^+ e^-$~\cite{Walcher:1970vkv}). The $N_*$ states are labeled by their energy above the nuclear ground state $N_0$ in MeV.  The $^8$Be excited states mix; we list the dominant isospin component. 
\vspace*{-.05in}
\label{table:states}}
\begin{tabular}[t]{ccccc}
 \toprule 
\rule{0pt}{0.4cm} $N_*$ & \ \quad $J_*^{P_*}$ \quad \ & \ \quad $T_*$ \quad \ & \ \quad $\Gamma_{N_*}$ (keV) \quad \ & \ \quad $B(N_* \to N_0 \gamma)$ \quad \ \\
\hline
\rule{0pt}{0.4cm} $^8$Be(18.15) &  $1^+$ & 0 & 138 & $1.4 \times 10^{-5}$  \\
\rule{0pt}{0.4cm} $^8$Be(17.64) &  $1^+$ & 1 & 10.7 & $1.4 \times 10^{-3}$  \\
\hline
\rule{0pt}{0.4cm} $^{12}$C(17.23) &  $1^-$ & 1 & 1150 & $3.8 \times 10^{-5}$ \\
\hline
\rule{0pt}{0.4cm} $^4$He(21.01) &  $0^-$ & 0 & 840 & 0  \\
\rule{0pt}{0.4cm} $^4$He(20.21) &  $0^+$ & 0 & 500  & \quad $6.6 \times 10^{-10}$ (E0)  \quad \\
\toprule
 \end{tabular}
\end{table}

For resonance production, the required proton energy is
\begin{equation}
E_p = \frac{m_{N_*}^2 - m_A^2 - m_p^2}{2 m_A} \ ,
\end{equation}
where $m_A$ and $m_{N_*}$ are the nuclear (not atomic) masses.  The proton beam energy, or proton kinetic energy, is $E_{\text{beam}} = E_p - m_p$, the proton's momentum is $p_p = \sqrt{E_p^2 - m_p^2}$, and the $N_*$ nucleus is created with velocity $v_{N_*} = p_p/(E_p+m_A)$ in the lab frame.  The values of $E_{\text{beam}}$, $m_A$, $m_{N_*}$, and $v_{N_*}$ for the considered processes are given in \tableref{kinematics}, and the properties of the relevant excited nuclear states are summarized in \tableref{states}.  

The anomaly observed in the $^8$Be system is unambiguously associated with the $^8$Be(18.15) state~\cite{Krasznahorkay:2015iga}.  The $^8$Be(18.15) and $^8$Be(17.64) resonances do not overlap significantly, and the anomaly was found to appear and disappear as the proton beam energy was varied to scan through the $^8$Be(18.15) resonance.  

\begin{figure} [tbp]
\centering
\includegraphics[width=0.6\textwidth]{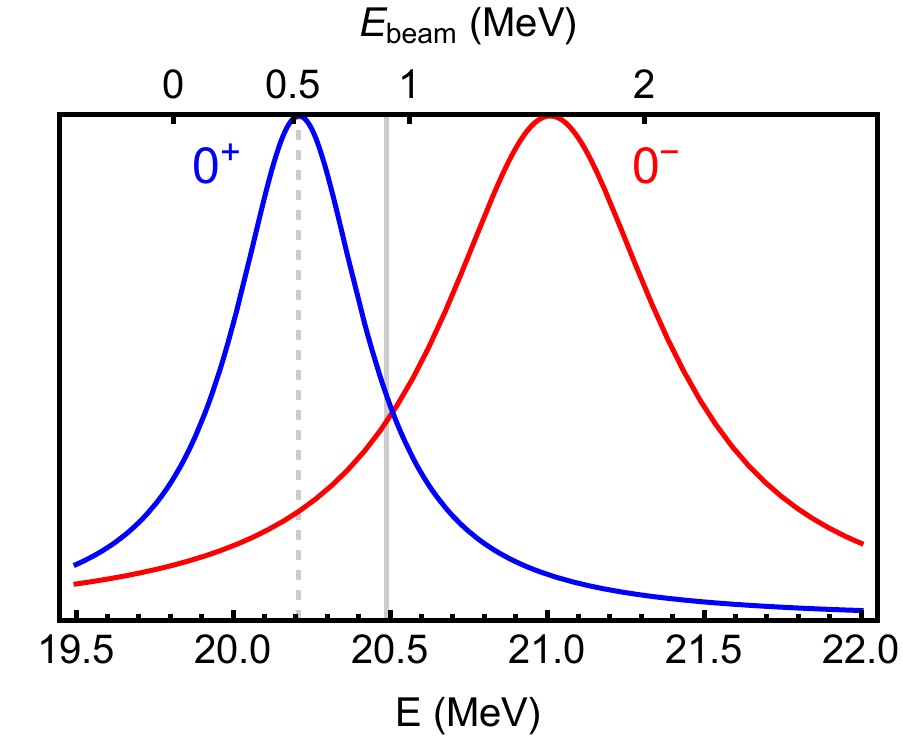} 
\caption{The Breit-Wigner resonance curves of the $0^+$ $^4$He(20.21) and $0^-$ $^4$He(21.01) excited states as a function of the excitation energy $E$ above the helium ground state (bottom) and the proton beam energy $E_{\text{beam}}$ (top). The solid and dashed vertical lines at beam energies of 0.90 MeV and 0.52 MeV correspond to the beam energy used in Ref.~\cite{Krasznahorkay:2019lyl} and the beam energy required to maximize the $0^{+}$ resonance, respectively.
\label{fig:BreitWig}}
\end{figure}

For the $^4$He anomaly~\cite{Krasznahorkay:2019lyl}, the experiment ran between the broad $0^+$ $^4$He(20.21) and $0^-$ $^4$He(21.01) resonances.  The proton beam energy was tuned to produce ${}^4$He nuclei with an excitation energy of $E =20.49$ MeV, which is well within the widths of both excited states; see \figref{BreitWig}.

\subsection{$X$ Boson Decay}
\label{sec:decay}

Once produced, the excited nucleus decays through $N_* \to N_0 X$.  In the $N_*$ rest frame, the $X$ boson is produced with energy 
\begin{equation}
E_X = \frac{m_{N_*}^2 + m_X^2 - m_{N_0}^2}{2 m_{N_*}} \approx m_{N_*} - m_{N_0} \ ,\label{eq:Exdef}
\end{equation}
where the last expression exploits the fact that $m_X, m_{N_*} - m_{N_0} \ll m_{N_*}, m_{N_0}$.  The $X$ velocity is $v_X = \sqrt{1 - (m_X/E_X)^2}$, and is shown in \tableref{kinematics} for the various processes, assuming $m_X = 17~\mev$.  The $X$ boson decays through $X \to e^+ e^-$.  In the $X$ rest frame, the electron and positron velocities are $v_e \approx 0.998$.

All of the processes of interest have the hierarchies:
\begin{equation}
v_{N_*} \alt 0.01 \ll v_X < v_e \approx 1 \ .
\end{equation}
Since $v_{N_*} \ll v_X$, $N_*$ can, to good approximation, be treated as being produced at rest in the lab frame.  In addition, because $v_X < v_e \approx 1$, the electron mass is negligible, and we may take $v_e$ to be 1.

With these simplifications, it is easy to determine the distribution of the opening angle $\theta_{e^+ e^-}$ in the lab frame.  In the $X$ rest frame, the electron and positron are produced back-to-back, with 4-momenta 
\begin{eqnarray}
p_{e^-}^X &=& \frac{m_X}{2} \, [1, \sin \theta, 0, \cos \theta] \label{eq:elecP}\\
p_{e^+}^X &=& \frac{m_X}{2} \, [1, -\sin \theta, 0, -\cos \theta] \, ,\label{eq:posP}
\end{eqnarray}
where $\theta$ is the angle relative to the $X$'s velocity in the lab frame.  Boosting in the $\hat{z}$ direction to the lab frame, these momenta become
\begin{eqnarray}
p_{e^-} &=&  \frac{m_X}{2}  \left[ \gamma_X (1+ v_X \cos \theta), \sin \theta, 0, \gamma_X (\cos \theta + v_X ) \right] \\
p_{e^+} &=&  \frac{m_X}{2}  \left[ \gamma_X (1- v_X \cos \theta), -\sin \theta, 0, \gamma_X (-\cos \theta + v_X ) \right] ,
\end{eqnarray}
where $\gamma_X = E_X/m_X$.  

\begin{figure} [tbp]
\centering
\includegraphics[width=0.7\textwidth]{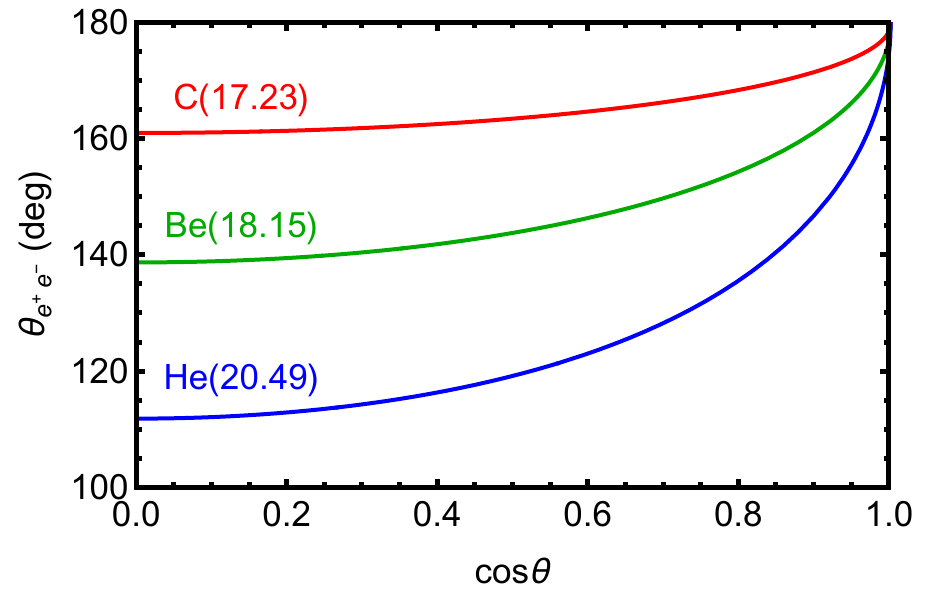} 
\caption{Opening angle $\theta_{e^+e^-}$ as a function of $\cos\theta$, where $\theta$ is the angle between the $e^{\pm}$ axis in the $X$ rest frame and the direction of the $X$ velocity in the lab frame.  Results are shown for decays from the excited nuclei indicated, assuming $m_X = 17~\mev$.  The opening angles are larger for smaller mass splittings.  Given a uniform distribution of $\cos\theta$, the opening angle distributions will be strongly peaked near their minimal values $\theta_{e^+e^-}^{\text{min}} = 161^\circ, 139^\circ$, and $112^\circ$ for the decays of $^{12}$C(17.23), $^8$Be(18.15), and $^4$He(20.49), respectively.}
\label{fig:thetadist}
\end{figure}

The opening angle is $\theta_{e^+ e^-} = \theta_{e^-} - \theta_{e^+}$, where 
\begin{eqnarray}
\theta_{e^-} &=& \tan^{-1} \! \left[ \frac{\sin \theta}{\gamma_X (\cos \theta + v_X )} \right] \quad {\rm and} \quad
\theta_{e^+} = \tan^{-1} \! \left[ \frac{-\sin \theta}{\gamma_X (-\cos \theta + v_X )} \right] , 
\end{eqnarray}
along with the appropriate choice of quadrant for $\theta_{e^\pm}$.  If $\cos \theta = 1$, the boost is (anti-)parallel to the $e^+/e^-$ direction in the $X$ rest frame, and the $e^-$ and $e^+$ remain back-to-back in the lab frame, so $\theta_{e^+e^-} = 180^\circ$.  On the other hand, if $\cos\theta = 0$, the boost is perpendicular to the $e^+/e^-$ direction in the $X$ rest frame, and the $e^-$ and $e^+$ velocities are both bent toward the boost direction an equal amount, yielding the minimal opening angle:
\begin{equation}
\theta_{e^-} = \tan^{-1} \! \left( \frac{m_X}{p_X} \right)  = - \theta_{e^+} \Rightarrow \theta_{e^+e^-}^{\text{min}} = 2 \sin^{-1} \! \left( \frac{m_X}{E_X} \right) 
\approx 2 \sin^{-1} \! \left( \frac{m_X}{m_{N_*} - m_{N_0}} \right)  .
\label{eq:thetamin}
\end{equation}
The opening angle $\theta_{e^+e^-}$ as a function of $\cos \theta$ is shown in \figref{thetadist}.  The possible opening angles range from $\theta_{e^+e^-}^{\text{min}}$ to $180^{\circ}$, as expected.  

However, if the $X$ decays are uniformly distributed in $\cos\theta$ (as they would be for spin-0 $X$ bosons), the distribution of opening angles will be highly peaked at $\theta_{e^+e^-}^{\text{min}}$.  For spin-1 $X$ bosons, the distribution could be modified by dynamical dependence on the $X$ spin direction, but it typically remains strongly peaked at the minimum value due to phase space.  

\begin{figure} [tbp]
\centering
\includegraphics[width=0.7\textwidth]{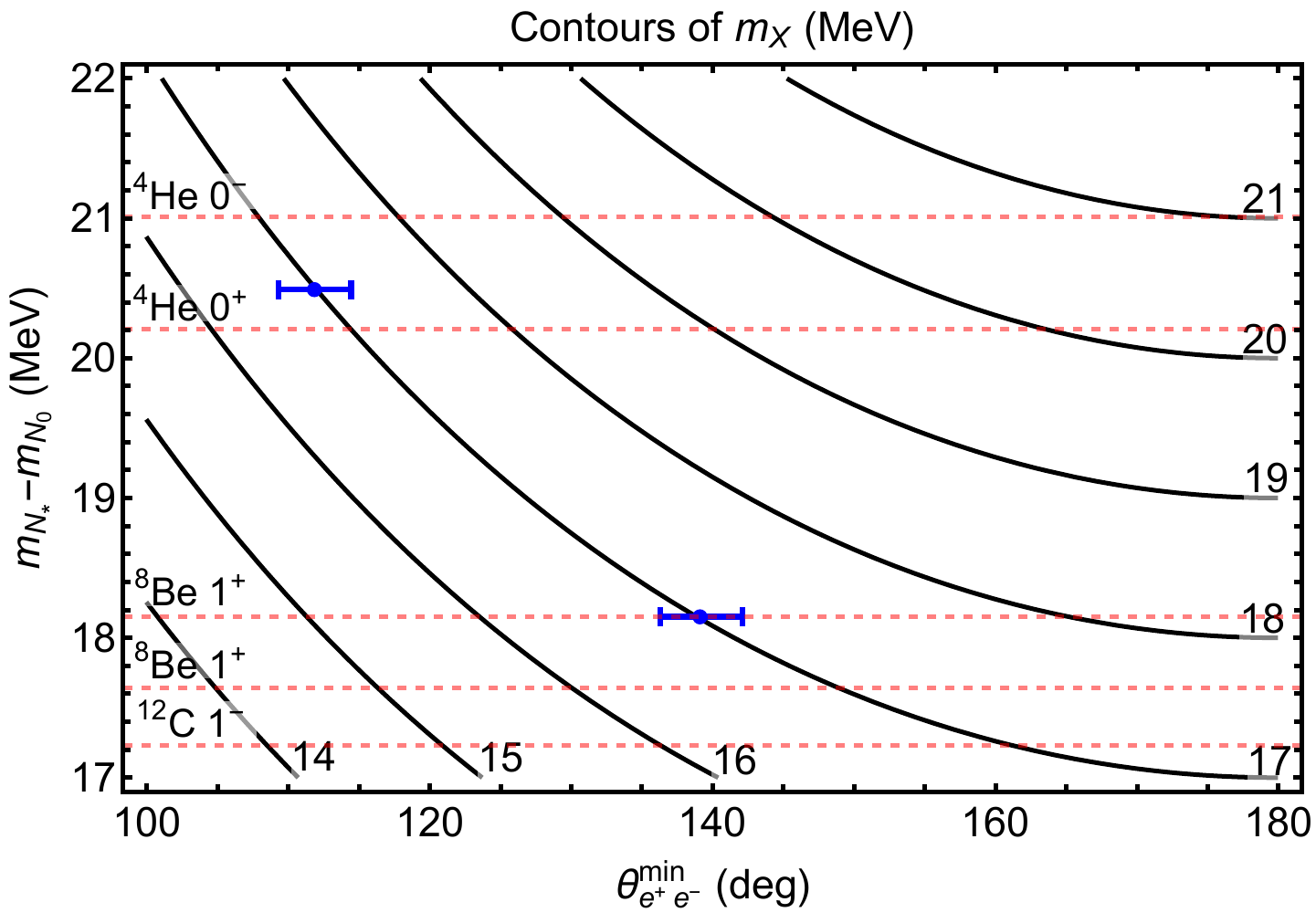} 
\caption{Contours (black solid lines) of the $X$ boson mass $m_X$ in the plane of the minimum opening angle $\theta_{e^+e^-}^{\text{min}}$ and the nuclear state mass 
splitting $m_{N_*} - m_{N_0}$. The relevant nuclear decay mass splittings discussed in the text are indicated by red dashed lines. 
The blue points and error bars indicate the parameters where $7\sigma$ excesses have been found in the opening angle distributions in 
$^8$Be and $^4$He nuclear decays. The excesses are at different opening angles $\theta_{e^+e^-}^{\text{min}}$, but both are consistent with the production of 17 MeV $X$ bosons.}
\label{fig:OpenAng}
\end{figure}

Thus, any $X$ bosons produced in the processes under consideration lead to an excess at $\theta_{e^+e^-}^{\text{min}}$. In \figref{OpenAng} we plot contours of $m_X$ in the $(\theta_{e^+e^-}^{\text{min}}, m_{N_*} - m_{N_0})$ plane, using the simple relation of \eqref{thetamin}.  The blue data points indicate the parameters where the $7\sigma$ excesses have been found in $^8$Be and $^4$He nuclear decays. The observations are at different opening angles, but both are consistent with the production of a boson with mass $m_X \approx 17~\mev$, a striking consistency check of the new particle hypothesis.

\section{Spin-Parity Analysis}
\label{sec:spinparity}

A first layer of particle dynamics, added to the kinematic foundation laid in the previous section, is to consider the allowed spin and parity assignments of an $X$ boson capable of accounting for the ATOMKI observations.  The assumption that parity is conserved warrants some discussion.  Although parity is a very good symmetry with respect to the internal nuclear dynamics and the electromagnetic interaction, the fundamental description of the $X$ boson interactions must be organized with respect to chirality and would require engineering to end up parity-symmetric.  At the same time, nuclear processes for the same spin, but opposite parity, typically proceed via different partial waves, and since the $X$ is produced at most semi-relativistically (see \tableref{kinematics}), the decay of a mixed state is likely to be highly dominated by a single parity component.  With this in mind, we organize our discussion by treating the $X$ boson as a state of definite parity, and consider the more general case of a vector particle with mixed vector and axial vector interactions in \secref{general}.

We continue to label each nuclear decay process as $N_* \to N_0 X$, where $N_*$ is an excited nucleus and $N_0$ is the ground state nucleus.  The cases of primary interest are those where anomalies have been observed: $N = {}^8$Be~\cite{Krasznahorkay:2015iga},  where the excited states are $1^+$ states, and $^4$He~\cite{Krasznahorkay:2019lyl}, where the excited states are $0^+$ and $0^-$ states.   However, to complete the analysis of all $J^P$ combinations up to spin 1, we also consider the case $N = \, ^{12}$C, where $^{12}$C(17.23) is an excited $1^-$ state that can be used to search for the  $X$ boson (see, e.g., Ref.~\cite{deBoer:2005kf}).  

The spin-parity of $N_*$, $N_0$, and $X$ are denoted by $J_*^{P_*}$, $J_0^{P_0}$, and $J_X^{ P_X}$, respectively.  Parity and angular momentum conservation imply
\begin{eqnarray}
J_* &=& L \oplus J_0 \oplus J_X \\
P_* &=& (-1)^L P_0 \, P_X \ ,
\end{eqnarray}
where $L$ is the final state orbital angular momentum, and $\oplus$ denotes the addition of angular momentum. In all the cases considered, the ground states have quantum numbers $J_0^{P_0} = 0^+$, leading to 
\begin{eqnarray}
J_* &=& L \oplus J_X \\
P_* &=& (-1)^L P_X \ .
\end{eqnarray}
We analyze the implications of these conservation laws for $J_X^{ P_X}$, considering spin-1 nuclear excited states and spin-0 excited states in turn.  

\subsection{Spin-1 Excited States: $^8$Be and $^{12}$C}

In the case of $^8$Be, the excited states of interest are both $1^+$ states: one with energy 18.15 MeV and the other with energy 17.64 MeV.  The conservation laws for production from the $^8$Be(18.15) state require
\begin{eqnarray}
1 &=& L \oplus J_X \\
+1 &=& (-1)^L P_X \ .
\end{eqnarray}
If $J_X = 0$, then $L = 1$, and $P_X = -1$:  $X$ cannot be a scalar, but it can be a pseudoscalar produced in the $P$-wave.  If $J_X = 1$, then $L = 0, 1, 2$, and $P_X = +1, -1, +1$, respectively: $X$ can either be a vector produced in a $P$-wave, or an axial vector produced in an $S$- or $D$-wave. 

In the case of $^{12}$C, the excited state is a $1^-$ state with energy 17.23 MeV above the stable ground state.  Relative to the $^8$Be case, the parity is reversed, and so all the $^8$Be results apply with the substitutions scalar $\leftrightarrow$ pseudoscalar and vector $\leftrightarrow$ axial vector.  

\subsection{Spin-0 Excited States: $^4$He}

The $^4$He excited states of interest are the $0^-$ $^4$He(21.01) state and the $0^+$ $^4$He(20.21) state.  The anomaly is seen at an intermediate energy chosen to excite both resonances.  For the $0^-$ excited state, the conservation laws require
\begin{eqnarray}
0 &=&  L \oplus J_X \\
-1 &=& (-1)^L P_X \ .
\end{eqnarray}
If $J_X = 0$, then $L = 0$, and $P_X = -1$:  $X$ cannot be a scalar, but it can be a pseudoscalar produced in the $S$-wave. If $J_X = 1$, then $L = 1$, and $P_X = +1$:  $X$ cannot be a vector, but it can be an axial vector produced in a $P$-wave state.  Note that $X$ cannot be a vector gauge boson, and so symmetries also forbid decays to single photons for this transition.

For the $0^+$ excited state, the parity is reversed, and so all the $0^-$ results apply with the substitutions scalar $ \leftrightarrow$ pseudoscalar and vector $ \leftrightarrow$ axial vector.

\subsection{Spin-Parity Summary}

The spin-parity analysis results are summarized in \tableref{decayoperators}. All four $J_*^{P_*}$ possibilities up to spin 1 have been considered, and so these results are broadly applicable to many other nuclear decays.

\begin{table}[tbp]
 \caption{Nuclear excited states $N_*$, their spin-parity $J_*^{P_*}$, and the possibilities for $X$ (scalar, pseudoscalar, vector, axial vector) allowed by angular momentum and parity conservation, along with the operators that mediate the decay and references to the equation numbers where these operators are defined.  The operator subscripts label the operator's dimension and the partial wave of the decay, and the superscript labels the $X$ spin.  For example, $\mathcal{O}_{4P}^{(0)}$ is a dimension-4 operator that mediates a $P$-wave decay to a spin-0 $X$ boson. 
\vspace*{-0.05in}
\label{table:decayoperators} }
\begin{tabular}[t]{|cc|c|c|c|c|}
\hline
\rule[-0.2cm]{0pt}{0.7cm} $N_*$ &  $J_*^{P_*}$ \ & \quad Scalar $X$ \ \qquad & Pseudoscalar $X$ 
& Vector $X$ & Axial Vector $X$ \\
\hline
\rule[-0.2cm]{0pt}{0.7cm} $^8$Be(18.15) &  $1^+$ & --- & $\mathcal{O}_{4P}^{(0)}$~(\ref{eq:O04P}) & $\mathcal{O}_{5P}^{(1)}$~(\ref{eq:O15P}) & \ $\mathcal{O}_{3S}^{(1)}$ (\ref{eq:O13S}), $\mathcal{O}_{5D}^{(1)}$ (\ref{eq:O15D}) \ \\
\hline
\rule[-0.2cm]{0pt}{0.7cm} $^{12}$C(17.23) & $1^-$ & $\mathcal{O}_{4P}^{(0)}$~(\ref{eq:O04P}) & ---
& \ $\mathcal{O}_{3S}^{(1)}$~(\ref{eq:O13S}), $\mathcal{O}_{5D}^{(1)}$~(\ref{eq:O15D}) \ & $\mathcal{O}_{5P}^{(1)}$~(\ref{eq:O15P}) \\
\hline
\rule[-0.2cm]{0pt}{0.7cm} $^4$He(21.01) & $0^-$ & --- & $\mathcal{O}_{3S}^{(0)}$~(\ref{eq:O03S}) & --- & $\mathcal{O}_{4P}^{(1)}$~(\ref{eq:O14P}) \\
\hline
\rule[-0.2cm]{0pt}{0.7cm} $^4$He(20.21) & $0^+$ & $\mathcal{O}_{3S}^{(0)}$~(\ref{eq:O03S}) & --- & $\mathcal{O}_{4P}^{(1)}$~(\ref{eq:O14P}) & --- \\
\hline
 \end{tabular}
\end{table}

By comparing results from more than one excited nucleus, one can in principle determine whether two signals can be attributed to a single $X$ boson, and constrain the possible spin-parity assignments of the $X$ boson.  The current situation is somewhat clouded by the experimental decision to target a non-resonant $^4$He state, where the anomaly can be attributed to transition through either the $0^+$ or the $0^-$ intermediate state (or both, if parity is not conserved).  

That said, it is worth noting that if the $X$ particle is being produced in both $0^-$ $^4$He and $^8$Be decays, $X$ must be either a pseudoscalar or an axial vector; it cannot be a vector (in contrast to the conclusion of Ref.~\cite{Krasznahorkay:2019lyl}).  In addition, this case implies that the $^8$Be decay and the $^4$He decay are either $S$-wave and $P$-wave, or vice versa. On the other hand, if the $X$ particle is produced in both $0^+$ $^4$He and $^8$Be decays, the only possibility is that the $X$ boson is a vector and that both decays are $P$-wave.  We construct an effective field theory for these nuclear decays in \secref{OpAnalysis}, and then use it to examine the pseudoscalar, axial vector, and vector possibilities in more detail in Secs.~\ref{sec:pseudoscalar}, \ref{sec:axialvector}, and \ref{sec:VecX}, respectively.

\section{EFT for Nuclear Transitions\label{sec:OpAnalysis}}

The next refinement in exploring the dynamical consistency of the $^8$Be and $^4$He anomalies is to explore whether a single set of $X$ interactions can consistently account for both of them. This is complicated by the fact that a fundamental description of the $X$ boson specifies its coupling to quarks and leptons, whereas the states participating in the reaction are complicated nuclei.  It is useful to employ the language of an effective field theory, which exploits the fact that the typical momentum transfer for the transitions of interest are far smaller than the sizes of the participating nuclei, allowing for an expansion in $p \times r_N$, where $p$ denotes the typical energy/momentum of the transition, and $r_N \sim$~(200 MeV)$^{-1}$ characterizes the size of the nucleus.  

In this limit, each energy level of the nucleus can be represented as a separate point-like quantum field, with interactions described by terms in an effective Lagrangian,
\beq
{\mathcal L}_{I} = \sum_i c_i \, {\mathcal O}_i \ .
\eeq
The ${\mathcal O}_i$ are operators built out of combinations of fields with powers of the nuclear scale $\Lambda$ included to make them dimension 4, and the dimensionless Wilson coefficients $c_i$ encode the nuclear physics.  Corrections due to finite size effects are represented by (even) higher dimensional non-renormalizable interactions.  The coefficients describing the interactions can in principle be extracted from theoretical nuclear physics computations or matched to experimental observations of nuclear transitions.  In cases where neither is available, they must be estimated using dimensional analysis.  As above, we take the spin-1 and spin-0 cases in turn.  In each case we denote the $0^+$ ground state by $N_0$ and the excited state by $N_\ast^{(\mu)}$, where a vector index is included when it is a spin-1 state. 

The EFT consists of all possible interaction terms consistent with the symmetries, including Lorentz invariance, gauge invariance (for the photon), baryon number, and (to good approximation) parity and strong isospin.  We review its construction in detail, though in some cases the material is well-known.  Rather than break the operators into components to analyze parity, we recognize that the contraction of a vector $V^\mu$ ($J^P =1^{-}$) with a derivative is parity even,
\beq
V^\mu\partial_\mu \xrightarrow{\text{Parity}} V^\mu\partial_\mu \ ,
\eeq
while the contraction of a derivative with an axial vector $A^\mu$ ($J^P = 1^+$) is parity odd,
\beq
A^\mu\partial_\mu \xrightarrow{\text{Parity}} -A^\mu\partial_\mu \ .
\eeq
Similarly, the contraction of two vector or two axial vectors is parity even, while the contraction of a vector with an axial vector is odd. These observations account for nearly all of the needed operators. The remaining cases employ the Levi-Civita tensor $\varepsilon^{\mu\nu\alpha\beta}$. If an operator that includes four contracted vector indices has definite parity, the operator composed of the same fields but with the vector indices contracted into the Levi-Civita tensor has the opposite parity.

We also make use of the equations of motion for a massive vector:
\beq
\left(\square+m_V^2 \right)V^\mu=J^\mu_V \ , \quad \partial_\mu V^\mu=0 \ ,
\eeq
where $m_V$ and $J_V^\mu$ are the mass and current of the $V^\mu$ particle, respectively.  Unlike in the massless case, the second condition is unrelated to a choice of gauge and implies that, in general, $p_\mu \epsilon_V^\mu(\vec{p})=0$, where $\epsilon_V^\mu(\vec{p})$ is the polarization vector of a $V^\mu$ state with momentum $p^\mu=(\sqrt{m_V^2+\vec{p}^{\,2}}\,,\vec{p}\,)$. This can be rewritten as
\beq
\epsilon^0_V(\vec{p}_V)=\frac{p^i_V}{E_V} \, \epsilon^i_V(\vec{p}_V)\ .\label{eq:polVec}
\eeq
Note that for a massive vector at rest, $\epsilon^0_V(\vec{0})=0$ for all three polarization states.

Finally, it is well known that simply enumerating all operators consistent with a given set of symmetries tends to introduce ambiguities in how the dynamics is described, although the physical outputs are the same. This is because operators are often related by integration by parts or through the equations of motion for a given field. In what follows we choose to focus on the operators that make agreement with the partial wave analysis in the decay process most transparent. In doing so, derivatives are typically moved from $N_0$ to $X$, but the final results do not depend on this choice. 

\subsection{Spin-1 Excited States: $^8$Be and $^{12}$C \label{sec:BeOp}}

In the spin-1 cases of beryllium and carbon, the excited state $N_\ast^\mu$ transforms as $N_\ast^\mu\partial_\mu \xrightarrow{\text{Parity}} (-)N_\ast^\mu\partial_\mu$ when the state is a vector (axial vector). The lowest dimension operator describing the decay into a spinless $X$ particle is the dimension-4 operator
\beq
\mathcal{O}_{4P}^{(0)}=N_0^\dag \, N_\ast^\mu \, \partial_\mu X \ . \label{eq:O04P}
\eeq
Here and below, the operator's subscripts denote the operator's dimension and partial wave of the decay it mediates, and the superscript denotes the spin of the $X$ boson.  Operator $\mathcal{O}_{4P}^{(0)}$ is one of a family of operators that includes $(\partial_\mu N_0)^\dag N_\ast^\mu X$ and $N_0^\dag (\partial_\mu  N_\ast^\mu) X$.  All three of these operators are related to one another by integration by parts, and so only two are linearly independent.  However, $\partial_\mu N_\ast^\mu$ vanishes for on-shell processes by the equations of motion, so we may choose $\mathcal{O}_{4P}^{(0)}$ as the single independent operator describing processes mediated by this interaction for which $N_\ast^\mu$ is on-shell.

Since $N_0$ is parity even, we find that for the entire operator to be parity even when $N^\mu_\ast$ is a vector (axial vector), $X$ must be a scalar (pseudoscalar). To see what $N_\ast^\mu$ decay partial waves result from this operator, we consider the matrix element $\langle N_0 X|\mathcal{O}_{4P}^{(0)}| N_\ast\rangle$ in the $N_\ast^\mu$ rest frame, which takes the form
\beq
\mathcal{M}\propto \epsilon_\ast(\vec{0})^i \, p_{Xi} \ .
\eeq
Because of the single factor of $\vec{p}_X$, this operator mediates a $P$-wave decay, in agreement with the spin-parity analysis of the previous section. 

The leading operator for a massive spin-1 $X$ particle is
\beq
\mathcal{O}_{3S}^{(1)}=\Lambda \, N_0^\dag \, N_\ast^\mu \, X_\mu \ .  \label{eq:O13S}
\eeq
If $N_\ast^\mu$ is a vector (axial vector), then parity demands that $X^\mu$ must also be a vector (axial vector).  The $N_\ast^\mu$ decay amplitude is proportional to
\beq
\mathcal{M}\propto \Lambda \, \epsilon_\ast(\vec{0})^i \, \epsilon_{Xi}(\vec{p}_X) 
\label{eq:Spin1S}
\eeq
and is clearly $S$-wave, after summing over the $X$ polarization states. 

At higher order, there are three dimension-5 operators to consider. The first is
\beq
\mathcal{O}_{5S}^{(1)}=\frac{N_0^\dag}{\Lambda}\left(\partial_\mu N_\ast^\nu \right)\partial^\mu X_\nu \ .
\label{eq:O15S}
\eeq
The contraction of derivatives is parity even, requiring the contraction of $X^\mu$ with $N_\ast^\mu$ to be parity even as well. Thus, $X^\mu$ and $N_\ast^\mu$ must either both be vectors or both be axial vectors. By integrating by parts three times, this operator can be rewritten as
\begin{align}
\frac{N_0^\dag}{\Lambda}\left(\partial_\mu N_\ast^\nu \right)\partial^\mu X_\nu=&-\frac{N_{\ast \nu}}{\Lambda}\left( \partial_\mu N_0^\dag\right)\partial^\mu X^\nu-\frac{N_0^\dag}{\Lambda}N_\ast^\mu \square X_\mu\nonumber\\
=&\frac{X^\nu}{\Lambda}\left(\partial^\mu N_{\ast\nu} \right)\partial_\mu N_0^\dag-\frac{N_0^\dag}{\Lambda}N_\ast^\mu \square X_\mu+N_\ast^\mu X_\mu\frac{\square}{\Lambda}N_0^\dag \nonumber\\
=&-\frac{N_0^\dag}{\Lambda}\left(\partial_\mu N_\ast^\nu \right)\partial^\mu X_\nu-\frac{N_0^\dag}{\Lambda}N_\ast^\mu \square X_\mu+N_\ast^\mu X_\mu\frac{\square}{\Lambda}N_0^\dag-\frac{N_0^\dag}{\Lambda}X_\mu\square N_\ast^\mu ~.\label{eq:O5Seq}
\end{align}
Using the leading order equations of motion of the three fields, the final three terms are proportional to $\mathcal{O}_{3S}^{(1)}$, implying that this operator is not independent
\beq
\frac{N_0^\dag}{\Lambda}\left(\partial_\mu N_\ast^\nu \right)\partial^\mu X_\nu=-\frac{m_{N_\ast}^2-m_{N_0}^2+m_X^2}{2\Lambda}N_0^\dag X_\mu N_\ast^\mu =-\frac{m_{N_\ast}E_X}{\Lambda}N_0^\dag X_\mu N_\ast^\mu~,\label{eq:O5StoO3S}
\eeq
where we have used \eqref{Exdef} in the last equality. 

The next operator is
\beq
\mathcal{O}_{5D}^{(1)}=\frac{N_0^\dag}{\Lambda}\left(\partial_\mu N_\ast^\nu \right)\partial_\nu X^\mu \ ,
\label{eq:O15D}
\eeq
which again requires $X^\mu$ and $N^\mu_\ast$ to have equal parity. In the $N_\ast^\mu$ rest frame, where $p_\ast^\mu=(m_{N_\ast},0,0,0)$, the decay matrix element takes the form
\beq
\mathcal{M}\propto\frac{m_{N_\ast}}{\Lambda}\delta^0_\mu \, \epsilon_\ast(\vec{0})^i \, p_{Xi} \, \epsilon_X^\mu(\vec{p}_X)
=\frac{m_{N_\ast}}{\Lambda E_X} \, \epsilon_\ast(\vec{0})^i \, p_{Xi} \, p^j_{X} \, \epsilon_X^j(\vec{p}_X) \ ,
\label{eq:Spin1D}
\eeq
where we have used \eqref{polVec}. Since this depends on two factors of $\vec{p}_X$, it describes a $D$-wave decay. The factor of $m_{N_\ast}$ in the numerator of this decay might cause one to wonder if it is really suppressed relative to the $S$-wave decay. However, comparing \eqref{O5StoO3S} and \eqref{Spin1D}, we see that the ratio of the $D$-wave operator to the $S$-wave operator is parametrically
\beq
\frac{p_X^2}{E_X^2}=v_X^2 \ ,\label{eq:DwaveSup}
\eeq
which is exactly what we expect when we compare a $D$-wave amplitude with an $S$-wave amplitude. 

The final dimension-5 operator contracts the four vector indices of the previous two operators with the Levi-Civita tensor: 
\beq
\mathcal{O}_{5P}^{(1)}=\frac{N_0^\dag}{\Lambda}\varepsilon^{\mu\nu\alpha\beta}\left(\partial_\mu N_{\ast \nu} \right)\partial_\alpha X_\beta \ .
\label{eq:O15P}
\eeq
As mentioned in the introduction to this section, this implies that $X^\mu$ and $N^\mu_\ast$ have opposite parity, in contrast to the two previous operators, where they had equal parity.   In the rest frame of the excited state, the amplitude for this decay is proportional to
\beq
\mathcal{M}\propto\frac{m_{N_\ast}}{\Lambda}\varepsilon^{0ijk} \, \epsilon_{\ast i}(\vec{0}) \, p_{Xj} \, \epsilon_{Xk}(\vec{p}_X) \ ,
\eeq
which is $P$-wave. Thus, we have explicitly confirmed that the operator-based analysis conforms with the spin-parity analysis for both the beryllium and carbon decays in every particular.  These results are summarized in \tableref{decayoperators}.

\subsection{Spin-0 Excited States: $^4$He\label{sec:HeOp}}

For helium the operator analysis is nearly identical for both the scalar and pseudoscalar excited states. If $X$ is spin 0, the leading operator that mediates the decay is
\beq
\mathcal{O}_{3S}^{(0)}=\Lambda N_0^\dag N_\ast X \ .
\label{eq:O03S}
\eeq
To preserve parity, either both $N_\ast$ and $X$ are scalars or both are pseudoscalars and it is clear that the decay mediated by this operator is $S$-wave.

If $X^\mu$ is spin 1, the leading independent operator is
\beq
\mathcal{O}_{4P}^{(1)}=N_0^\dag X^\mu\partial_\mu N_\ast \ .
\label{eq:O14P}
\eeq
A similar operator with the derivative acting on $N_0$ is obtained by integrating by parts and recalling that $\partial_\mu X^\mu=0$. The parity analysis is simple: if $N_\ast$ is parity even, then so is $X^\mu \partial_\mu$, so $X^\mu$ is a vector. If $N_\ast$ is a pseudoscalar, then $X^\mu$ must be an axial vector. The amplitude for $N_\ast$ decay in its own rest frame is
\beq
\mathcal{M}\propto\epsilon_X^\mu(\vec{p}_X) \, m_{N_\ast}\delta_\mu^0 =\frac{m_{N_\ast}}{E_X} \, p^i_{X} \, \epsilon_X^i(\vec{p}_X) \ ,
\eeq
where we have used the identity in \eqref{polVec}. This is a $P$-wave decay, and again we find complete agreement between the spin-parity and operator analyses of the helium decays. These results are summarized in \tableref{decayoperators}.

\section{Pseudoscalar $X$}
\label{sec:pseudoscalar}

When $X$ is a pseudoscalar, its couplings to protons and neutrons can be written as
\beq
\mathcal{L}_X\supset X\left[\varepsilon_p\overline{p}\gamma_5p+\varepsilon_n\overline{n}\gamma_5n \right]
= X\left[ \frac12 (\varepsilon_p+\varepsilon_n) J^5_0
+ \frac12(\varepsilon_p-\varepsilon_n) J^5_1 \right] ,
\eeq
where
\begin{equation}
J^5_0 = \overline{p}\gamma_5p+\overline{n}\gamma_5n \quad \text{and} \quad
J^5_1 = \overline{p}\gamma_5p-\overline{n}\gamma_5n
\end{equation}
are the isosinglet and isotriplet currents, respectively. Nuclear transitions between states of the same isospin couple to $J^5_0$, while transitions that change isospin by one unit couple to $J^5_1$. 

For the $^8$Be case, the decay of the ${}^8$Be(18.15) excited nucleus to the ground state transitions from (predominantly) isosinglet to isosinglet, so $X$ must couple through the $J^5_0$ field.  In terms of nucleon currents, then, the decay amplitude is
\beq
\mathcal{M}=\langle N_0X|\frac12\left(\varepsilon_p+\varepsilon_n \right)XJ_0^5|N_\ast\rangle
= \frac12\left(\varepsilon_p+\varepsilon_n \right)\langle {}^8\text{Be} | J^5_0 | {}^8\text{Be}(18.15) \rangle \ .
\label{eq:Benucleon}
\eeq
In the nuclear EFT, the $P$-wave decay amplitude, determined by $\mathcal{O}^{(0)}_{4P}$ in \eqref{O04P}, can be written as 
\beq
\mathcal{M} = \langle N_0X|  \frac12\left(\varepsilon_p+\varepsilon_n \right) C_{P, \text{Be}} \, \mathcal{O}^{(0)}_{4P}|N_\ast\rangle
= \frac12\left(\varepsilon_p+\varepsilon_n \right) C_{P, \text{Be}}  \, \epsilon_\ast^\mu \, p_{X\mu} \ ,
\label{eq:Benuclear}
\eeq
where $C_{P, \text{Be}} $ is the Wilson coefficient of the operator. This coefficient is expected to be ${\cal{O}}(1)$ in the nuclear EFT~\cite{Manohar:1983md}.  From matching \eqsref{Benucleon}{Benuclear}, it is determined by
\begin{equation}
C_{P, \text{Be}}  \, \epsilon_\ast^\mu \, p_{X\mu} = \langle {}^8\text{Be} | J^5_0 | {}^8\text{Be}(18.15) \rangle \ ,
\label{eq:Bematrixelement}
\end{equation}
where the nuclear matrix element has been estimated in Ref.~\cite{Ellwanger:2016wfe} using a shell model. Note that, since the excited beryllium state is spin-1, the matrix element must depend on $\epsilon_\ast^\mu$. Because the matrix element is a scalar, this polarization vector must be contracted into some combination of the particle momenta. Conservation of momentum allows us to write this as a linear combination of $p_X^\mu$ and $p_\ast^\mu$, but the latter vanishes when contracted with $\epsilon_\ast^\mu$, leaving only $p_{X\mu}$ in \eqref{Bematrixelement}. Note also that the combination $\varepsilon_{\mu\nu\alpha\beta}\epsilon_{\ast}^\mu p_\ast^\nu p_0^\alpha p_X^\beta$ vanishes because the momenta are not independent, but could not appear in any case because it is odd under parity. 

The decay width is then
\beq
\Gamma^{{}^8\text{Be}}_P (N_\ast\to N_0 X)
= \frac{(\varepsilon_p+\varepsilon_n)^2 \, C_{P, \text{Be}}^2}{96\pi m_{N_\ast}^2} \, p_{X, \text{Be}}^3 \ ,
\label{eq:PBedecay}
\eeq
where $p_{X, \text{Be}}$ is the magnitude of the 3-momentum of the $X$ particle produced in the beryllium decay.  

For the $^{12}$C case, as shown in \secref{spinparity}, the $^{12}$C(17.23) $1^-$ state cannot decay into the $0^+$ ground state and a pseudoscalar.  If an $X$ boson were observed in $^{12}$C(17.23) decay, it would exclude the possibility that $X$ is a pure pseudoscalar.

In the ${}^4$He case, the decay is from the $^4$\text{He}(21.01) $0^-$ isosinglet state.  In terms of nucleon currents, the decay amplitude is therefore
\beq
\mathcal{M}=\langle N_0X|\frac12\left(\varepsilon_p+\varepsilon_n \right)XJ_0^5|N_\ast\rangle
= \frac12\left(\varepsilon_p+\varepsilon_n \right)\langle {}^4\text{He} | J^5_0 | {}^4\text{He}(21.01) \rangle \ .
\label{eq:Henucleon}
\eeq
In the nuclear EFT, the decay is mediated by the operator $\mathcal{O}_{3S}^{(0)}$ in \eqref{O03S}, leading to the $S$-wave decay amplitude
\beq
\mathcal{M} = \langle N_0X|  \frac12\left(\varepsilon_p+\varepsilon_n \right) C_{P, \text{He}} \, \mathcal{O}^{(0)}_{3S}|N_\ast\rangle
= \frac12\left(\varepsilon_p+\varepsilon_n \right) C_{P, \text{He}}  \, \Lambda  \ .
\eeq
The ${\cal{O}}(1)$ Wilson coefficient $C_{P, \text{He}}$ is determined simply by
\begin{equation}
C_{P, \text{He}} \, \Lambda = \langle {}^4\text{He} | J^5_0 | {}^4\text{He}(21.01) \rangle \ ,
\end{equation}
and the decay width is
\beq
\Gamma^{{}^4\text{He}}_P (N_\ast\to N_0 X)=\frac{(\varepsilon_p+\varepsilon_n)^2 \, C_{P, \text{He}}^2 \, 
\Lambda^2}{32\pi m_{N_\ast}^2}p_{X,\text{He}} \ .
\label{eq:PHedecay}
\eeq

Therefore, for pseudoscalar $X$ bosons, the ratio of the beryllium and helium decay widths of \eqsref{PBedecay}{PHedecay} is predicted to be 
\beq
\frac{\Gamma_P^{{}^8\text{Be}}}{\Gamma_P^{{}^4\text{He}}}
=  \frac13 \, \frac{C_{P, \text{Be}}^2}{C_{P, \text{He}}^2} \, \frac{m_{\text{He}_*}^2}{m_{\text{Be}_*}^2} \, 
\frac{p_{X,\text{Be}}}{p_{X,\text{He}}} \, \frac{p_{X,\text{Be}}^2}{\Lambda^2}
\approx 1.7 \times10^{-6} \ \frac{C_{P, \text{Be}}^2}{C_{P, \text{He}}^2} \left( \frac{\text{GeV}}{\Lambda}\right)^2 ,
\eeq
where we have assumed $m_X=17$ MeV.  The coefficients $C_{P, \text{Be}}$ and $C_{P, \text{He}}$ are expected to be ${\cal{O}}(1)$ in the EFT, but they are not both readily available in the literature, leading to significant uncertainty in their ratio.  On the experimental side, the fact that the ATOMKI ${}^4$He data were collected off-resonance requires some care (see \secref{general}). Nevertheless, even given these theoretical and experimental uncertainties, the EFT prediction $\Gamma_P^{{}^8\text{Be}} /\Gamma_P^{{}^4\text{He}} \sim 10^{-6}$ is clearly difficult to reconcile with the observations of \eqsref{Bewidth}{width}, which imply $\Gamma_P^{{}^8\text{Be}} \sim \Gamma_P^{{}^4\text{He}}$, and a pseudoscalar $X$ explanation for both anomalies is strongly disfavored.
 
Note also that the extreme hierarchy in expected decay widths also implies that $X$ is unlikely to be a scalar and pseudoscalar mixture, since a significant pseudoscalar component is required to explain the beryllium anomaly (since the beryllium state cannot decay to a scalar), but even a small pseudoscalar component would result in a very large helium decay width, which has not been observed.

\section{Axial Vector $X$}
\label{sec:axialvector}

Similar to the pseudoscalar case, an axial vector $X$ couples to the nucleon currents
\beq
J_X^{\mu 5} = \varepsilon_p \overline{p}\gamma^\mu\gamma_5p + \varepsilon_n  \overline{n}\gamma^\mu\gamma_5n 
=\frac12 (\varepsilon_p + \varepsilon_n) J_0^{\mu 5} + \frac{1}{2} (\varepsilon_p - \varepsilon_n) J_1^{\mu 5}  \ ,
\eeq
where 
\begin{equation}
J^{\mu 5}_0 = \overline{p}\gamma^\mu \gamma_5 p+\overline{n} \gamma^\mu \gamma_5 n \quad \text{and} \quad
J^{\mu 5}_1 = \overline{p}\gamma^\mu \gamma_5 p-\overline{n} \gamma^\mu \gamma_5 n
\end{equation}
are the isospin-preserving and isospin-changing currents, respectively. 

For the isospin-preserving ${}^8$Be decay, the nucleon-level amplitude is
\beq
\mathcal{M}=\langle N_0X|\frac12\left(\varepsilon_p+\varepsilon_n \right)X_\mu J_0^{\mu 5}|N_\ast\rangle=\frac12\left(\varepsilon_p+\varepsilon_n \right) \langle {}^8\text{Be} | J^{\mu 5}_0 | {}^8\text{Be}(18.15) \rangle \epsilon_{X\mu} \ .
\eeq
At the nucleus level, the decay is mediated by the operator $\mathcal{O}^{(1)}_{3S}$ in \eqref{O13S}, which leads to the $S$-wave decay amplitude
\beq
\mathcal{M}=\langle N_0X| \frac12\left(\varepsilon_p+\varepsilon_n \right) C_{A, \text{Be}} \, \mathcal{O}^{(1)}_{3S}|N_\ast\rangle= \frac12\left(\varepsilon_p+\varepsilon_n \right) C_{A, \text{Be}} \, \Lambda \, \epsilon_\ast^\mu \, \epsilon_{X\mu} \ ,
\eeq
where $C_{A, \text{Be}}$ is the $\mathcal{O}(1)$ Wilson coefficient of the operator. Matching these amplitudes, we find
\begin{equation}
C_{A, \text{Be}} \, \Lambda \, \epsilon_\ast^\mu = \langle {}^8\text{Be} | J^{\mu 5}_0 | {}^8\text{Be}(18.15) \rangle \ ,
\end{equation}
where the nuclear matrix element has been obtained via many-body techniques in Ref.~\cite{Kozaczuk:2016nma}.  Note that the nuclear matrix element must be proportional to $\epsilon_\ast^\mu$. The quantity $\varepsilon^{\mu\nu\alpha\beta} \epsilon_{\ast \nu} p_{\ast \alpha} p_{X \beta}$ cannot appear because it would make the matrix element odd under parity.  The resulting decay width is
\beq
\Gamma^{{}^8\text{Be}}_A (N_\ast\to N_0 X)=
\frac{(\varepsilon_p+\varepsilon_n)^2 \, C_{A, \text{Be}}^2 \, \Lambda^2}{32\pi m_{N_\ast}^2} \, p_{X, \text{Be}} \left(1+\frac{p_{X, \text{Be}}^2}{3m_X^2} \right).
\label{eq:AVBeWidth}
\eeq

For ${}^{12}$C, the decay of the excited nucleus requires changing isospin by one unit, and so the nucleon-level amplitude is 
\beq
\mathcal{M}=\langle N_0X|\frac12\left(\varepsilon_p-\varepsilon_n \right)X_\mu J_1^{\mu 5}|N_\ast\rangle=\frac12\left(\varepsilon_p-\varepsilon_n \right) \langle {}^{12}\text{C} | J^{\mu 5}_1 | {}^{12}\text{C}(17.23) \rangle \epsilon_{X\mu} \ .
\eeq
In the nuclear EFT, the operator $\mathcal{O}^{(1)}_{5P}$ of \eqref{O15P} gives rise to the amplitude
\beq
\mathcal{M}=\langle N_0X| \frac12\left(\varepsilon_p-\varepsilon_n \right) C_{A, \text{C}} \, \mathcal{O}^{(1)}_{5P}|N_\ast\rangle
= \frac12\left(\varepsilon_p-\varepsilon_n \right) C_{A, \text{C}} \, \frac{1}{ \Lambda} \varepsilon^{\mu\nu\alpha\beta} \, p_{\ast \mu} \, \epsilon_{\ast \nu} \, p_{X \alpha} \, \epsilon_{X \beta} \ ,
\eeq
where the Wilson coefficient $C_{A, \text{C}}$ is determined by
\begin{equation}
C_{A, \text{C}} \, \frac{1}{\Lambda} \varepsilon^{\mu\nu\alpha\beta} \, p_{\ast\nu} \, \epsilon_{\ast\alpha} \, p_{X\beta}
= \langle {}^{12}\text{C} | J^{\mu 5}_1 | {}^{12}\text{C}(17.23) \rangle \ .
\end{equation}
The excited nuclear state has spin 1, so the nuclear matrix element must be proportional to $\epsilon_\ast^\mu$. However, because $N_\ast^\mu$ is a vector, while $X_\mu$ is an axial vector, the inner product of their polarization vectors is parity odd. Therefore, the matrix element must be proportional to the axial vector $\varepsilon^{\mu\nu\alpha\beta} p_{\ast\nu}\epsilon_{\ast\alpha} p_{X\beta}$. The resulting decay width is
\beq
\Gamma^{{}^{12}\text{C}}_A (N_\ast\to N_0 X)=\left(\varepsilon_p-\varepsilon_n \right)^2\frac{C_{A, \text{C}}^2}{48\pi\Lambda^2} \, p_{X, {\rm C}}^3 \ .
\label{eq:AVCWidth}
\eeq

Combining \eqsref{AVBeWidth}{AVCWidth}, the ratio of the beryllium and carbon decay widths is
\begin{align}
\frac{\Gamma_A^{{}^8\text{Be}}}{\Gamma_A^{{}^{12}\text{C}}} = 
& \ \frac32 \frac{\left(\varepsilon_p+\varepsilon_n \right)^2}{\left(\varepsilon_p-\varepsilon_n \right)^2} \,
\frac{C_{A, \text{Be}}^2}{C_{A, \text{C}}^2} \,
\frac{\Lambda^4}{m_{\text{Be}_\ast}^2} \frac{p_{X, \text{Be}}}{p_{X, \text{C}}^3} 
\left(1+\frac{p_{X, \text{Be}}^2}{3m_X^2} \right) \nonumber\\
\approx & \ 8.1\times 10^{3} \ \frac{\left(\varepsilon_p+\varepsilon_n \right)^2}{\left(\varepsilon_p-\varepsilon_n \right)^2} \, \frac{C_{A, \text{Be}}^2}{C_{A, \text{C}}^2} \left( \frac{\Lambda}{\text{GeV}}\right)^4 .
\end{align}
Although this result is quantitatively limited by our knowledge of $C_{A, \text{Be}}$ and $C_{A, \text{C}}$, the se coefficients are expected to be $\mathcal{O}(1)$ in the nuclear EFT, and so we expect $\Gamma_A^{^8\text{Be}} \gg \Gamma_A^{^{12}\text{C}}$.  Given that the carbon total decay width and photon decay width are both about an order of magnitude larger than their beryllium counterparts, for pseudoscalar $X$ bosons, the carbon decay rate will be greatly suppressed relative to the beryllium decay rate and very difficult to observe. 

The ${}^4$He decay must proceed from the $^4$\text{He}(21.01) state.  It is isospin-preserving, and so the nucleon-level amplitude is
\beq
\mathcal{M}=\langle N_0X|\frac12\left(\varepsilon_p+\varepsilon_n \right)X_\mu J^{\mu 5}_0|N_\ast\rangle= \frac12\left(\varepsilon_p+\varepsilon_n \right) \langle {}^4\text{He} | J^{\mu 5}_0 | {}^4\text{He}(21.01) \rangle \,
\epsilon_{X\mu}\ .
\eeq
At the nucleus level, the $P$-wave decay is mediated by $\mathcal{O}^{(1)}_{4P}$ in \eqref{O14P}, with amplitude 
\beq
\mathcal{M}=\langle N_0X| \frac12\left(\varepsilon_p+\varepsilon_n \right) C_{A, \text{He}} \, \mathcal{O}^{(1)}_{4P}|N_\ast\rangle
= \frac12\left(\varepsilon_p+\varepsilon_n \right) C_{A, \text{He}} \, p_\ast^\mu \, \epsilon_{X\mu} \ ,
\eeq
where 
\begin{equation}
C_{A, \text{He}} \, p_\ast^\mu  = P_{X\nu}^\mu \langle {}^4\text{He} | J^{5\nu}_0 | {}^4\text{He}(21.01) \rangle \ ,
\end{equation}
and
\beq
P_{X\nu}^\mu=\delta^\mu_\nu-\frac{p_X^\mu\,p_{X\nu}}{p_X^2} \label{eq:Xproject}
\eeq
is the projection matrix into the subspace orthogonal to $p_X^\mu$.  The matrix element has a vector Lorentz index, but is composed of scalar fields. This, along with momentum conservation, implies that $\langle {}^4\text{He} | J^{\mu 5}_0 | {}^4\text{He}(21.01) \rangle$ is proportional to a linear combination $p_\ast^\mu$ and $p_X^\mu$. However, the latter vanishes when contracted with the $X$ polarization vector.  The decay width is 
\beq
\Gamma^{{}^4\text{He}}_A (N_\ast\to N_0 X)
=\frac{(\varepsilon_p+\varepsilon_n)^2 \, C_{A, \text{He}}^2 }{32\pi m_X^2} \, p_{X, \text{He}}^3 \ .
\label{eq:AVHeWidth}
\eeq

Thus, combining \eqsref{AVBeWidth}{AVHeWidth}, for an axial vector $X$ boson, we find
\beq
\frac{\Gamma_A^{{}^8\text{Be}}}{\Gamma_A^{{}^4\text{He}}}
= \frac{C_{A,\text{Be}}^2}{C_{A,\text{He}}^2}  \, \Lambda^2 \, \frac{m_X^2}{m_{\text{Be}_\ast}^2} \, \frac{p_{X, \text{Be}}}{p_{X, \text{He}}^3} \left(1+\frac{p_{X, \text{Be}}^2}{3m_X^2} \right) 
\approx 1.8 \times10^{-2} \ \frac{C_{A,\text{Be}}^2}{C_{A,\text{He}}^2} \left( \frac{\Lambda}{\text{GeV}}\right)^2 ,
\eeq
where we have assumed $m_X=17$ MeV.  In contrast to the pseudoscalar case, the decay widths are not expected to be separated by two orders of magnitude, not four.  Within the uncertainties of the measurements and the nuclear matrix elements, it may be possible that an axial vector $X$ boson could explain both the $^8$Be and $^4$He anomalies.

\section{Vector $X$ \label{sec:VecX}}

In contrast to the pseudoscalar and axial vector cases, the necessary nuclear matrix elements for decays to a pure $J^P = 1^-$ vector $X$ boson are related to standard model decays to photons and E0 transitions and cancel out in appropriately defined ratios.  

\subsection{Beryllium Decays to Vector $X$ Bosons}

We begin by reviewing the analysis for beryllium decays~\cite{Feng:2016jff,Feng:2016ysn}.  The vector $X$ boson couples to the nucleon current 
\beq
J_X^{\mu} = \varepsilon_p \, e \, \overline{p}\gamma^\mu p + \varepsilon_n \, e \, \overline{n}\gamma^\mu n 
=\frac12 (\varepsilon_p + \varepsilon_n) e \, J_0^{\mu} + \frac{1}{2} (\varepsilon_p - \varepsilon_n) e \, J_1^{\mu}  \ ,
\eeq
where 
\begin{equation}
J^{\mu}_0 = \overline{p}\gamma^\mu p+\overline{n}\gamma^\mu n \quad \text{and} \quad
J^{\mu}_1 = \overline{p}\gamma^\mu p-\overline{n}\gamma^\mu n \label{eq:isoJs}
\end{equation}
are the isospin-preserving and isospin-changing vector currents, respectively. In contrast to the previous two sections, we have included a factor of the photon coupling $e$ to simplify comparisons between $X^\mu$ and photon decays. 

$^8$Be(18.15) and the $^8$Be ground state have equal isospin.  If we assume isospin is conserved and neglect isospin mixing in the nuclear states, only the $J_0^{\mu}$ current contributes, and the decay amplitude is 
\begin{eqnarray}
\mathcal{M} =\langle {}^8\text{Be} \,X| X_\mu J_X^{\mu} | {}^8\text{Be}(18.15) \rangle &=&\frac{1}{2} (\varepsilon_p + \varepsilon_n)\, e \, \langle {}^8\text{Be} | J_0^{\mu} | {}^8\text{Be}(18.15) \rangle \epsilon_{X\mu} \ .
\end{eqnarray}
At the nucleus level, the $P$-wave decay is mediated by $\mathcal{O}^{(1)}_{5P}$ in \eqref{O15P}, leading to the decay amplitude
\beq
\mathcal{M}=\langle N_0 X | \frac{1}{2} (\varepsilon_p + \varepsilon_n) \, e \, C_{V, \text{Be}} \, \mathcal{O}^{(1)}_{5P}|N_\ast\rangle= \frac{1}{2} (\varepsilon_p + \varepsilon_n) \, e \, C_{V, \text{Be}} \frac{1}{\Lambda} 
\varepsilon^{\mu\nu\alpha\beta} \, p_{\ast\mu} \, \epsilon_{\ast\nu} \, p_{X\alpha} \, \epsilon_{X\beta} \ .
\eeq
Matching the amplitudes, we find that the Wilson coefficient $C_{V, \text{Be}}$ is determined by
\begin{equation}
C_{V,\text{Be}} \, \frac{1}{\Lambda} \varepsilon^{\mu\nu\alpha\beta} \, p_{\ast\nu} \, \epsilon_{\ast\alpha} \, p_{X \beta} 
= \langle {}^8\text{Be} | J_0^{\mu} | {}^8\text{Be}(18.15) \rangle \ .
\end{equation}
Because the excited nuclear state is spin-1, the nuclear matrix element must contain a factor of $\epsilon_\ast^\mu$. However, since $N_\ast^\mu$ is an axial vector, while $X^\mu$ is a vector, the matrix element must be proportional to $\varepsilon^{\mu\nu\alpha\beta} p_{\ast\nu}\,\epsilon_{\ast\alpha}\,p_{X \beta} $ to preserve parity. 

The resulting decay width is
\beq
\Gamma^{^8\text{Be}}_X (N_\ast\to N_0 X)= (\varepsilon_p + \varepsilon_n)^2 \, \frac{\alpha \, C_{V,\text{Be}}^2}{12 \Lambda^2} \, p_{X, \text{Be}}^3 \ ,
\eeq
where $p_{X,\text{Be}}$ is the magnitude of the 3-momentum of the $X$ boson.  For the photon, the width is similarly computed.  Note that the mediating operator is ``accidentally'' gauge invariant, $\mathcal{O}_{5P}^{(1)}= 1/(2 \Lambda) N_0^\dag \varepsilon^{\mu\nu\alpha\beta}\left(\partial_\mu N_{\ast \nu} \right) X_{\alpha\beta}$, where $X_{\alpha\beta}$ is the $X$ boson's field strength, and so the results have a smooth limit $m_X \to 0$ and are immediately applicable to photons.  The resulting photon width is
\beq
\Gamma^{^8\text{Be}}_\gamma (N_\ast\to N_0 \gamma )= \frac{\alpha \, C_{V,\text{Be}}^2}{12 \Lambda^2} \, p_{\gamma, \text{Be}}^3 \ ,
\eeq
and the ratio of widths is
\beq
\frac{\Gamma^{{}^8\text{Be}}_X}{\Gamma^{{}^8\text{Be}}_\gamma}
=(\varepsilon_p+\varepsilon_n)^2~\frac{p_{X,\text{Be}}^3}{p_{\gamma, \text{Be}}^3}\approx(\varepsilon_p+\varepsilon_n)^2\left[1-\frac{m_X^2}{(m_{N_\ast}-m_{N_0})^2} \right]^{3/2} \ .
\eeq
Because the same nuclear matrix element occurs in both the vector $X$ boson and photon decays, the ratio of the two decay widths is particularly simple and independent of nuclear matrix elements.

As shown in Ref.~\cite{Feng:2016ysn}, the presence of the nearby spin-1, isospin-triplet $^8$Be(17.64) state induces some isospin mixing corrections to this result. Including the isospin mixing, the ratio of widths is modified to
\begin{eqnarray}
\frac{\Gamma^{{}^8\text{Be}}_X}{\Gamma^{{}^8\text{Be}}_\gamma}&=&\left|-0.09 \, (\varepsilon_p+\varepsilon_n)+1.09 \, (\varepsilon_p-\varepsilon_n)\right|^2 \, \frac{p_{X,\text{Be}}^3}{p_{\gamma, \text{Be}}^3} \nonumber \\
&\approx &  0.043 \left|-0.09 \, (\varepsilon_p+\varepsilon_n)+1.09 \, (\varepsilon_p-\varepsilon_n)\right|^2 ,
\label{eq:isospinmixingwidth}
\end{eqnarray}
where we have set $m_X=17~\mev$ in deriving the final numerical result.  The coefficients of the isosinglet and isotriplet components are related to the degree of mixing and relative strengths of the respective M1 transitions in the ${}^8$Be system. A further refinement to fit experimental data by introducing isospin breaking in the electromagnetic transition operators was introduced in Ref.~\cite{Feng:2016ysn}.  With both isospin mixing and isospin breaking, the width ratio is modified to
\begin{eqnarray}
\frac{\Gamma^{{}^8\text{Be}}_X}{\Gamma^{{}^8\text{Be}}_\gamma}&=&\left| 0.05 \, (\varepsilon_p+\varepsilon_n)+ 0.95 \, (\varepsilon_p-\varepsilon_n)\right|^2 \, \frac{p_{X,\text{Be}}^3}{p_{\gamma, \text{Be}}^3} \nonumber \\
&\approx &  0.043 \left| 0.05 \, (\varepsilon_p+\varepsilon_n)+ 0.95 \, (\varepsilon_p-\varepsilon_n)\right|^2 .
\label{eq:isospinmixingbreakingwidth}
\end{eqnarray}
We use both \eqsref{isospinmixingwidth}{isospinmixingbreakingwidth} in presenting our results below. 

\subsection{Carbon Decays to Vector $X$ Bosons \label{sec:CarbonVector}}

For ${}^{12}$C, it is straightforward to calculate the $X$ boson decay rate. The excited state is a vector with isospin $T_* =1$. Consequently, the transition to the ground state $0^+$ is through the $J_1^\mu$ current of \eqref{isoJs}.  As with the case of $^8$Be decaying to axial vector $X$ bosons, this carbon decay is mediated by the operator $\mathcal{O}^{(1)}_{3S}$ of \eqref{O13S}, but now with both $N_\ast^\mu$ and $X^\mu$ vectors instead of both axial vectors.  The decay amplitudes are
\beq
\mathcal{M}=\langle N_0X|\frac12 (\varepsilon_p-\varepsilon_n ) \, e \, X_\mu J_1^{\mu}|N_\ast\rangle
= \frac12 (\varepsilon_p - \varepsilon_n ) \, e \, \langle {}^{12}\text{C} | J_1^{\mu} | {}^{12}\text{C}(17.23) \rangle \epsilon_{X\mu} 
\eeq
and
\beq
\mathcal{M}=\langle N_0X| \frac12 (\varepsilon_p-\varepsilon_n ) \, e \, C_{V, \text{C}} \, \mathcal{O}^{(1)}_{3S}|N_\ast\rangle= \frac12 (\varepsilon_p-\varepsilon_n ) \, e \, C_{V, \text{C}} \, \Lambda \, \epsilon_\ast^\mu \, \epsilon_{X\mu} \ ,
\eeq
where 
\begin{equation}
C_{V,\text{C}} \, \Lambda\,\epsilon^\mu_\ast =  \langle {}^{12}\text{C} | J_1^{\mu} | {}^{12}\text{C}(17.23) \rangle \ ,
\end{equation}
and the resulting decay width is (see \eqref{AVBeWidth})
\beq
\Gamma^{^{12}\text{C}}_X (N_\ast\to N_0 X)
= (\varepsilon_p-\varepsilon_n)^2 \frac{\alpha \, C_{V,\text{C}}^2 \, \Lambda^2}{8 m_{N_\ast}^2}p_{X, \text{C}} \left(1+\frac{p_{X, \text{C}}^2}{3m_X^2} \right) .
\label{eq:VCWidth}
\eeq

As with beryllium, we can relate the $X$ decay rate to the photon decay rate and form a ratio that is independent of nuclear matrix elements. Unfortunately, the procedure is less straightforward than in the beryllium case, because, as is clear from the $m_X \to 0$ limit of \eqref{VCWidth}, the analysis above is not appropriate for the photon decay. This is simply because, when the photon is substituted for $X$ in $\mathcal{O}^{(1)}_{3S}$, the result is not gauge invariant.  Instead of using $\mathcal{O}^{(1)}_{3S}$ immediately, we must first determine the gauge-invariant operator that mediates photon decay and then relate it to $\mathcal{O}^{(1)}_{3S}$. The decay amplitude is
\beq
\mathcal{M}=\langle N_0X| \frac12 \, e \, C_{\gamma, \text{C}} \, \mathcal{O}_{\gamma}|N_\ast\rangle \ ,\eeq
where $\mathcal{O}_{\gamma}$ is the gauge-invariant operator, and $C_{\gamma, \text{C}}$ is its Wilson coefficient.  

At leading order, there appear to be two possible dimension-5 gauge-invariant operators: 
\beq
\mathcal{O}_{\gamma}=\frac{1}{\Lambda}F^{\mu\nu}N_0^\dag\partial_\nu N_{\ast\mu} 
\quad \text{and} \quad 
\mathcal{O}'_{\gamma} = \frac{1}{\Lambda}F^{\mu\nu}N_{\ast\mu}\partial_\nu N_0^\dag \ ,
\label{eq:photonoperator}
\eeq
where $F^{\mu \nu}$ is the usual electromagnetic field strength. However, integrating by parts, we find
\beq
\mathcal{O}_{\gamma} 
= - \mathcal{O}'_{\gamma} - \frac{1}{\Lambda}N_{\ast\mu}N_0^\dag\partial_\nu F^{\mu\nu} \ .
\eeq
Given the photon equation of motion $\partial_\mu F^{\mu\nu}=J_{\text{EM}}^\nu$, the last term is suppressed by the small coupling in the photon current. To leading order, then, the two operators are interchangeable, and the difference involving the current $J_{\text{EM}}^{\nu}$ can be incorporated systematically through Feynman diagrams. We can therefore simply consider $\mathcal{O}_{\gamma}$. 

To relate $\mathcal{O}_{\gamma}$ to $\mathcal{O}^{(1)}_{3S}$, we replace $F^{\mu\nu}\to X^{\mu\nu}$ in $\mathcal{O}_{\gamma}$ and use
\begin{eqnarray}
\frac{1}{\Lambda}X^{\mu\nu}N_0^\dag\partial_\nu N_{\ast\mu} &=& \frac{1}{\Lambda}\left(\partial^\mu X^\nu\right)N_0^\dag\partial_\nu N_{\ast\mu} -\frac{1}{\Lambda}\left(\partial^\nu X^\mu\right)N_0^\dag\partial_\nu N_{\ast\mu} \nonumber\\
&=&\frac{1}{\Lambda}\left(\partial^\mu X^\nu\right)N_0^\dag\partial_\nu N_{\ast\mu}+ \frac{m_{N_\ast}E_X}{\Lambda}N_0^\dag N_\ast^\mu X_\mu~,\label{eq:VecCarbonOp}
\end{eqnarray}
where, in the second line, we have applied the identity given in \eqref{O5StoO3S}.  The second term on the right has the same form as $\mathcal{O}^{(1)}_{3S}$ in \eqref{O13S}. The first mediates the $D$-wave decay, and is smaller than the dimension 3 term by a factor of $v_X^2\sim0.03$ (see \eqref{DwaveSup} and \tableref{kinematics}), so we neglect it. We find, then, that the Wilson coefficients of $\mathcal{O}^{(1)}_{3S}$ and $\mathcal{O}_\gamma$ are related by \
\beq
C_{\gamma,\text{C}}  = C_{V, \text{C}} \, \frac{ \Lambda^2}{m_{N_\ast}E_X} \ .\label{eq:CarbonLambda}
\eeq
With this identification, we can calculate the decay amplitude for the photon decay, and determine an expression for the ratio of $X$ boson and photon widths that is independent of nuclear matrix elements.  

It is instructive to determine the decay amplitude in a general way that encompasses both the massless and massive vector cases.  Specifically, we determine the amplitude
\beq
\mathcal{M}=\langle N_0 X | \frac12 \, \varepsilon \, e \, C_{\gamma, \text{C}} \, \mathcal{O}_V |N_\ast\rangle \ ,
\eeq
where
\beq
\mathcal{O}_V = \frac{\kappa}{\Lambda}\left(\partial^\mu V^\nu\right) N_0^\dag \, \partial_\nu N_{\ast\mu} 
-\frac{1}{\Lambda}\left(\partial^\nu V^\mu\right)N_0^\dag \, \partial_\nu N_{\ast\mu} \ .
\eeq
When $\kappa=1$, $\mathcal{O}_V$ is $\mathcal{O}_\gamma$, the gauge-invariant operator of \eqref{photonoperator} that mediates decays to a massless photon.  But when $\kappa=0$, $\mathcal{O}_V$ is proportional to $\mathcal{O}^{(1)}_{3S}$, the operator that mediates decays to a massive $X$, and we expect to recover the decay width calculated in \eqref{VCWidth}. The amplitude is
\beq
\mathcal{M}=\frac12 \, \varepsilon \, e \, C_{\gamma,\text{C}} \, \frac{1}{\Lambda} \, \epsilon_{\ast \mu} \, \epsilon_{V\nu}\left[\kappa\, p_V^\mu p_\ast^\nu-\left(p_V\cdot p_\ast\right)\eta^{\mu\nu} \right] ,
\eeq
which leads to the decay width
\beq
\Gamma_V^{^{12}\text{C}}(N_\ast\to N_0 V) = \varepsilon^2 \frac{\alpha \, C_{\gamma,\text{C}}^2}{24\Lambda^2} \, p_{V,\text{C}} \left[ 3m_V^2+2(2-\kappa) \, p_{V,\text{C}}^2+(1-\kappa)^2 \, \frac{p_{V,\text{C}}^4}{m_V^2}\right] .
\eeq

For the photon, we take $\kappa=1$, $m_V \to 0$, and $\varepsilon = 1$ to obtain
\beq
\Gamma_{\gamma}^{^{12}\text{C}}(N_\ast\to N_0 \gamma) = \frac{\alpha \, C_{\gamma,\text{C}}^2}{12\Lambda^2} \, p_{\gamma,\text{C}}^3 \ .
\label{eq:gammaCWidth}
\eeq
For the massive vector $X$ boson, we take $\kappa=0$ and $\varepsilon = \varepsilon_p - \varepsilon_n$ to find  
\beq
\Gamma_X^{^{12}\text{C}}(N_\ast\to N_0 X)= (\varepsilon_p-\varepsilon_n)^2 \, \frac{\alpha \, C_{\gamma,\text{C}}^2}{8\Lambda^2} \, p_{X,\text{C}} \, E_{X,\text{C}}^2 \left(1+\frac{p_{X,\text{C}}^2}{3m_X^2} \right).
\label{eq:XCWidth}
\eeq
Using \eqref{CarbonLambda} to write $C_{\gamma, \text{C}}$ in terms of $C_{V, \text{C}}$, we recover exactly the result of \eqref{VCWidth}, confirming the Wilson coefficient relation of \eqref{CarbonLambda}.

Combining \eqsref{gammaCWidth}{XCWidth} yields
\begin{eqnarray}
\frac{\Gamma^{^{12}\text{C}}_X}{\Gamma^{^{12}\text{C}}_\gamma} &=& \frac32 \, (\varepsilon_p-\varepsilon_n)^2 \, \frac{p_{X,\text{C}}}{p_{\gamma,\text{C}}^3} \, E_{X,\text{C}}^2 \left(1+\frac{p_{X,\text{C}}^2}{3m_X^2} \right) \approx 0.25 \  (\varepsilon_p-\varepsilon_n)^2 \ ,
\label{eq:Carbonwidth}
\end{eqnarray}
where we have set $m_X=17~\mev$ in deriving the final numerical result.  The ratio is independent of nuclear matrix elements, as desired, and the singularity as $m_X\to0$ is now readily understood. 

\subsection{Helium Decays to Vector $X$ Bosons}

For ${}^{4}$He, it is also straightforward to calculate the decay rate into the $X$ boson. The $0^-$ state cannot decay to vector bosons, but the $0^+$ state can.  The $0^+$ state has the same isospin as the ground state, and so the amplitudes are
\beq
\mathcal{M}=\langle N_0 X | \frac12 \, (\varepsilon_p + \varepsilon_n) e \, X_\mu J_0^\mu|N_\ast\rangle=\frac12 \, ( \varepsilon_p+\varepsilon_n ) \, e \, \langle{}^4\text{He}|J_0^\mu|{}^4\text{He}(20.21)\rangle \epsilon_{X\mu}  
\label{eq:HeXnucamp}
\eeq
and
\beq
\mathcal{M} = \langle N_0X|  \frac12 \, (\varepsilon_p + \varepsilon_n) \, e \, C_{V,\text{He}} \, \mathcal{O}_{4P}^{(1)}|N_\ast\rangle
=  \frac12 \, (\varepsilon_p + \varepsilon_n) \, e \, C_{V,\text{He}} \, p_\ast^\mu \, \epsilon_{X\mu} \ .
\label{eq:HeXamp}
\eeq
Matching these two amplitudes, the Wilson coefficient is determined by
\begin{equation}
C_{V,\text{He}} \, p_\ast^\mu = P^\mu_{X\nu}\langle{}^4\text{He}|J_0^\nu|{}^4\text{He}(20.21)\rangle \ , 
\end{equation}
where $P^\mu_{X\nu}$, defined in \eqref{Xproject}, projects into the subspace orthogonal to $p_X^\mu$. The nuclear states are scalars, so the Lorentz index in the matrix element must be attached to a linear combination of the particle momenta. By momentum conservation, this can be chose to be a combination of $p_\ast^\mu$ and $p_X^\mu$, but the latter vanishes when contracted with the $X$ polarization vector. The resulting decay rate is
\begin{equation}
\Gamma^{^4\text{He}}_X (N_* \to N_0 X) 
= (\varepsilon_p + \varepsilon_n)^2 \, \frac{\alpha \, C_{V,\text{He}}^2}{8 m_X^2} \, p_{X, \text{He}}^3 \ .
\label{eq:HeXWidth}
\end{equation}
 
Following the ${}^8$Be and ${}^{12}$C examples above, we hope to normalize the $X$ decay rate to the photon decay rate.  Unfortunately, the relevant $^4$He excited states do not decay to photons.  For the $0^-$ state, the spin-parity analysis of \secref{spinparity} showed that decays to vector bosons, either massless or massive, are forbidden.  This is also easily seen from the operator point of view. The only gauge-invariant operator that could mediate a photon decay is 
\beq
\frac{1}{\Lambda^2}\varepsilon^{\mu\nu\alpha\beta} \, F_{\mu\nu} \,  \partial_\alpha N_0^\dag \,  \partial_\beta N_\ast \ .
\eeq
Upon integration by parts, we find this operator vanishes, either because of the symmetry of the derivative operator or by the Bianchi equation, $\varepsilon^{\mu\nu\alpha\beta}\partial_\beta F_{\mu\nu}=0$. 

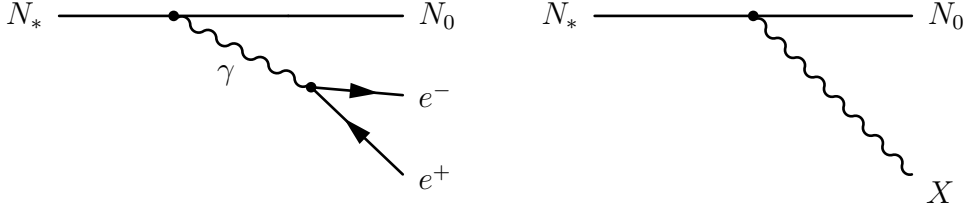
\begin{figure}[t]
\centering
\begin{fmffile}{E0}
\begin{fmfgraph*}(130,60)
\fmfpen{1.0}
\fmfstraight
\fmfleft{p1,p1,i1}\fmfright{o1,o2,o3}
\fmfv{l= $N_\ast$,l.a=180}{i1}\fmfv{l=$e^+$,l.a=-10}{o1}
\fmfv{l= $e^-$}{o2}\fmfv{l=$N_0$,l.a=0}{o3}
\fmf{plain,tension=1}{i1,v1,vp,o3}
\fmffreeze
\fmf{boson, tension=0.8,label=$\gamma$,label.side=right}{v1,v2}
\fmf{fermion,tension=0.6}{o1,v2,o2}
\fmfv{decor.shape=circle,decor.filled=full,decor.size=1.5thick}{v1}
\fmfv{decor.shape=circle,decor.filled=full,decor.size=1.5thick}{v2}
\end{fmfgraph*}
\end{fmffile}\hspace{2cm}
\begin{fmffile}{Xdecay}
\begin{fmfgraph*}(120,60)
\fmfpen{1.0}
\fmfstraight
\fmfleft{p1,i1}\fmfright{o2,o3}
\fmfv{l= $N_\ast$,l.a=180}{i1}
\fmfv{l= $X$}{o2}\fmfv{l=$N_0$,l.a=0}{o3}
\fmf{plain,tension=1}{i1,v1,o3}
\fmffreeze
\fmf{boson, tension=0.8}{v1,o2}
\fmfv{decor.shape=circle,decor.filled=full,decor.size=1.5thick}{v1}
\end{fmfgraph*}
\end{fmffile}
\caption{\label{fig:E0} Feynman diagrams for the E0 transition (left) and the decay into $X^\mu$ (right).}
\end{figure}

The $0^+$ state can decay to a massive vector boson, but unfortunately, it also cannot decay to a photon.  When $N_\ast$ is the $0^+$ state, we can write the gauge-invariant operator
\beq
\mathcal{O}_{\text{E0}}=\frac{1}{\Lambda^2}F^{\mu\nu} \, \partial_\mu N_0^\dag \, \partial_\nu N_\ast \ .
\label{eq:E0}
\eeq
However, if we use this operator to calculate the decay width $\Gamma (N_\ast\to N_0\gamma)$, we find the amplitude vanishes. This occurs because we can integrate $\mathcal{O}_{\text{E0}}$ by parts and use the photon equation of motion $\partial_\mu F^{\mu\nu}=0+J_{\text{EM}}^\nu$.  It is this leading zero that is captured by the amplitude, reflecting the fact that the $0^+$ excited state cannot decay into the ground state and an on-shell photon.  The piece that is higher order in the electromagnetic coupling,
\beq
\frac{1}{\Lambda^2}J_{\text{EM}}^\nu N_0^\dag \partial_\nu N_\ast \ ,
\label{eq:E0current}
\eeq
is non-vanishing and mediates the E0 transition $0^+ \to {}^4\text{He} \, e^+e^-$ shown in \figref{E0}. This is an electromagnetic transition, but one in which the photon is never on-shell. 

It is possible, however, to normalize the $X$ boson decay to the $0^+$ E0 transition.  Following the procedure of \secref{CarbonVector}, we relate $\mathcal{O}_{\text{E0}}$ to $\mathcal{O}_{4P}^{(1)}$.  The E0 decay amplitude is
\beq
\mathcal{M}=\langle N_0X| \frac12 \, e \, C_{\text{E0}} \, \mathcal{O}_{\text{E0}}|N_\ast\rangle \ ,
\eeq
where $C_{\text{E0}}$ is the Wilson coefficient of the $\mathcal{O}_{\text{E0}}$ operator. Similar decays to $e^+e^-$ from the $0^-$ state can only appear though operators of higher dimension and are consequently neglected. 

We can relate the $\mathcal{O}_\text{E0}$ operator to the massive vector operator $\mathcal{O}_{4P}^{(1)}$ in \eqref{O14P} by using the $X$ equations of motion. If we begin with the same operator as in \eqref{E0}, but replace $F^{\mu\nu}$ with the $X$ field strength $X^{\mu\nu}$, we have
\beq
\frac{1}{\Lambda^2}X^{\mu\nu} \, \partial_\mu N_0^\dag \, \partial_\nu N_\ast \ .
\eeq
Integrating by parts and using the $X$ equations of motion $\partial_\mu X^{\mu\nu}+m_X^2X^\nu=J_X^\nu$, we find this operator includes
\beq
\frac{m_X^2}{\Lambda^2}X^\mu N_0^\dag\partial_\mu N_\ast \ .
\label{eq:Ox}
\eeq
which is exactly the form of $\mathcal{O}_{4P}^{(1)}$ in \eqref{O14P}.  The remaining term that includes the $X^\mu$ current is subleading, because it contains additional factors of the small $X$ coupling, and its effects can be included diagrammatically. The Wilson coefficients of $\mathcal{O}^{(1)}_{4P}$ and $\mathcal{O}_\text{E0}$ are therefore related by
\beq
C_{\text{E0}} = \frac{\Lambda^2}{m_X^2} \, C_{V,\text{He}} \ .
\label{eq:HeliumLambda}
\eeq

Now that we have related the Wilson coefficients, we can calculate the E0 decay width. In Lorentz gauge, $\partial_\mu A^\mu=0$, and so the E0 operator of \eqref{E0} is 
\begin{align}
\mathcal{O}_\text{E0}=&\frac{1}{\Lambda^2}\left(\partial^\mu A^\nu-\partial^\nu A^\mu \right)\left( \partial_\mu N_\ast\right)\partial_\nu N_0^\dag\nonumber\\
=&-\frac{1}{\Lambda^2}\left[ \partial^\mu\partial_\nu A^\nu-\partial^2A^\mu\right]N_0^\dag\partial_\mu N_\ast\nonumber\\
=&\frac{1}{\Lambda^2}N_0^\dag\left(\partial_\mu N_\ast \right)\partial^2A^\mu \ ,
\end{align}
which makes clear that the interaction vanishes when the photon is on-shell. We use this vertex to calculate the $N_\ast$ decay shown in the left panel of \figref{E0}. Summing over final state spins, we find the squared amplitude 
\beq
\sum_\text{spin}\left|\mathcal{M} \right|^2=\frac{e^4 \,C_{\text{E0}}^2}{\Lambda^4}\left[2(p_+\cdot p_0)(p_-\cdot p_0)-m_{N_0}^2(p_+\cdot p_-)-m_e^2m_{N_0}^2 \right] ,
\eeq
where $p_{\pm}$ is the $e^{\pm}$ momentum, and $p_0$ is the $N_0$ momentum. 

The integration over the three-body phase space is most conveniently expressed as integrals over the electron and positron energies, $E_+$ and $E_-$,
\begin{align}
\Gamma_\text{E0}=&\frac{\alpha^2 \, C_{\text{E0}}^2 \, m_{N_\ast}}{8\pi \Lambda^4}\int dE_+ \,dE_-\left[ m_{N_\ast}^2-m_{N_0}^2-2m_{N_\ast}\left(E_++E_- \right)+4E_+E_- \right] ,
\end{align}
with the limits of integration chosen so that the outer (e.g., $E_+$) limits of integration are
\beq
E_{+\text{Min}}=m_e \ , \quad E_{+\text{Max}} = \frac{\left( m_{N_\ast}-m_e\right)^2-m_{N_0}^2+m_e^2}{2\left( m_{N_\ast}-m_e\right)} \ ,
\eeq
while the inner limits are
\begin{eqnarray}
E_{-(\pm)}&=&\frac{1}{2\left[m_{N_\ast}(m_{N_\ast}-2 E_+)+m_e^2 \right]}\left\{\left(m_{N_\ast}-E_+\right)\left[m_{N_\ast}\left(m_{N_\ast}-2 E_+\right)-m_{N_0}^2+2m_e^2 \right]\phantom{\sqrt{\left[m_{N_0}^2\right]}}\right.\nonumber\\
&&\left.\pm\sqrt{\left(m_{N_\ast}^2-E_+^2\right)\left[\left( m_{N_\ast}^2-2m_{N_\ast} E_+-m_{N_0}^2\right)^2-4m_{N_0}^2m_e^2\right]}\right\} ,
\end{eqnarray}
where the positive and negative signs correspond to the maximum and minimum, respectively. 

In general, the integral does not have a simple solution. However, neglecting the electron mass results in
\begin{eqnarray}
\Gamma_\text{E0}&\approx& \frac{\alpha^2 \, C_{\text{E0}}^2 \, m_{N_\ast}^5}{8\pi \Lambda^4}
\left[\frac{1}{96}\left( 1-8\frac{m_{N_0}^2}{m_{N_\ast}^2}+8\frac{m_{N_0}^6}{m_{N_\ast}^6}-\frac{m_{N_0}^8}{m_{N_\ast}^8}\right)+\frac{m_{N_0}^4}{4m_{N_\ast}^4}\ln\frac{m_{N_\ast}}{m_{N_0}} \right] \nonumber\\
&\approx& \frac{\alpha^2 \, C_{\text{E0}}^2 \, m_{N_\ast}^5}{8\pi \Lambda^4}\frac{2}{15}\left( \frac{m_{N_\ast}-m_{N_0}}{m_{N_0}}\right)^5 \nonumber\\
&\approx& \frac{\alpha^2 \, C_{\text{E0}}^2 \, m_{N_\ast}^5}{8\pi \Lambda^4} \left( 6.03\times 10^{-13} \right) \ .
\label{eq:HeE0Width}
\end{eqnarray}
Evaluating the integral numerically with the electron mass included, we find the final numerical coefficient changes by a tiny amount to $5.97\times 10^{-13}$.

Combining \eqsref{HeXWidth}{HeE0Width}, and using \eqref{HeliumLambda} to relate $C_{\text{E0}}$ and $C_{V,\text{He}}$, we find that the ratio of the $X$ width to the E0 width is 
\begin{eqnarray}
\frac{\Gamma^{^4\text{He}}_X}{\Gamma_\text{E0}} &\approx & \frac{\left(\varepsilon_p+\varepsilon_n \right)^2}{6.0 \times 10^{-13}} \, \frac{\pi}{\alpha} \, \frac{m_X^2\, p_{X,\text{He}}^3}{m_{N_\ast}^5} 
\approx 360 \left( \varepsilon_p+\varepsilon_n \right)^2  ,
\label{eq:Heliumwidth}
\end{eqnarray}
where $\Gamma_\text{E0}=(3.3\pm1)\times10^{-4}$ eV~\cite{Walcher:1970vkv}, and we have set $m_X = 17~\mev$ in deriving the final numerical result. The ratio is independent of nuclear matrix elements, and is larger than ratios of $X$ widths to photon widths derived here, since the E0 transition is a 3-body decay.

\subsection{Vector $X$ Summary}

For the case of a vector boson, the nuclear decay widths are functions of $m_X$ and the couplings $\varepsilon_p$ and $\varepsilon_n$.  The constraints on these parameters from all other experiments have been determined in Refs.~\cite{Feng:2016jff,Feng:2016ysn}.   A bound from NA48/2 on the process $\pi \to X \gamma$~\cite{Batley:2015lha} requires $|\varepsilon_p| < 1.2 \times 10^{-3}$, that is, that the $X$ boson be protophobic.  Once this constraint is satisfied, no other constraints further restrict the values of $m_X$ and $\varepsilon_n$ favored by the beryllium anomaly.  

\begin{figure} [tbp]
\centering
\includegraphics[width=0.43\textwidth]{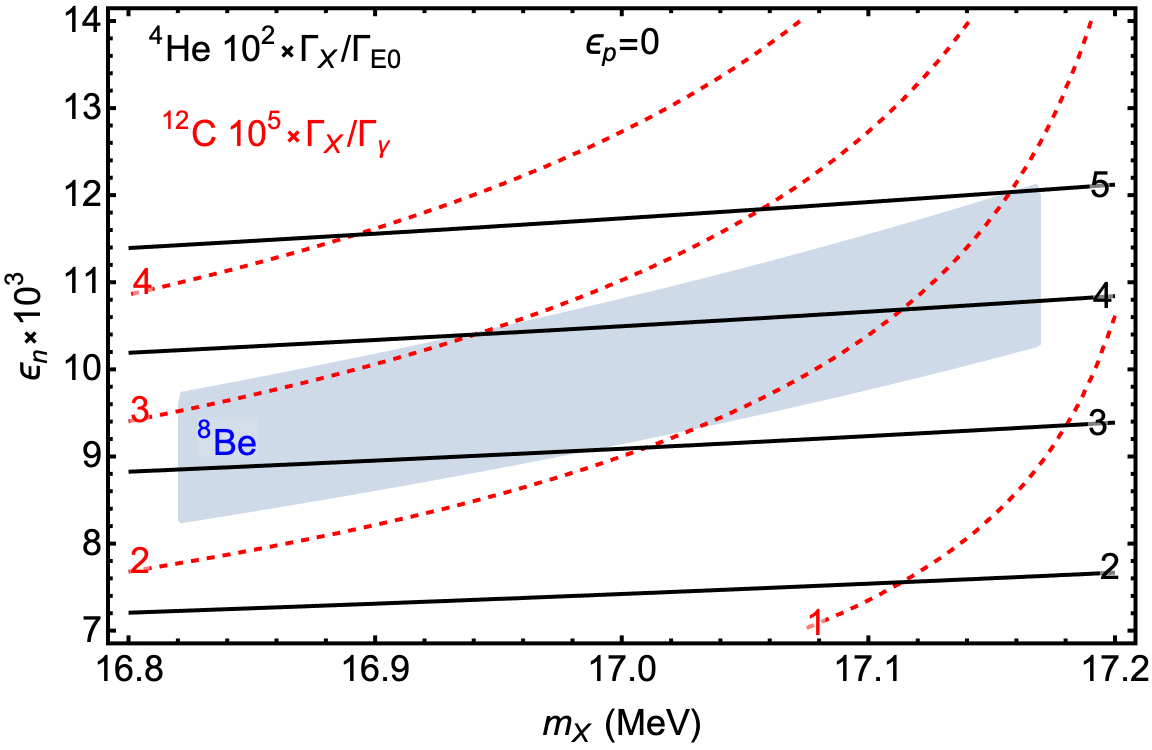} \\
\vspace*{0.1in}
\includegraphics[width=0.43\textwidth]{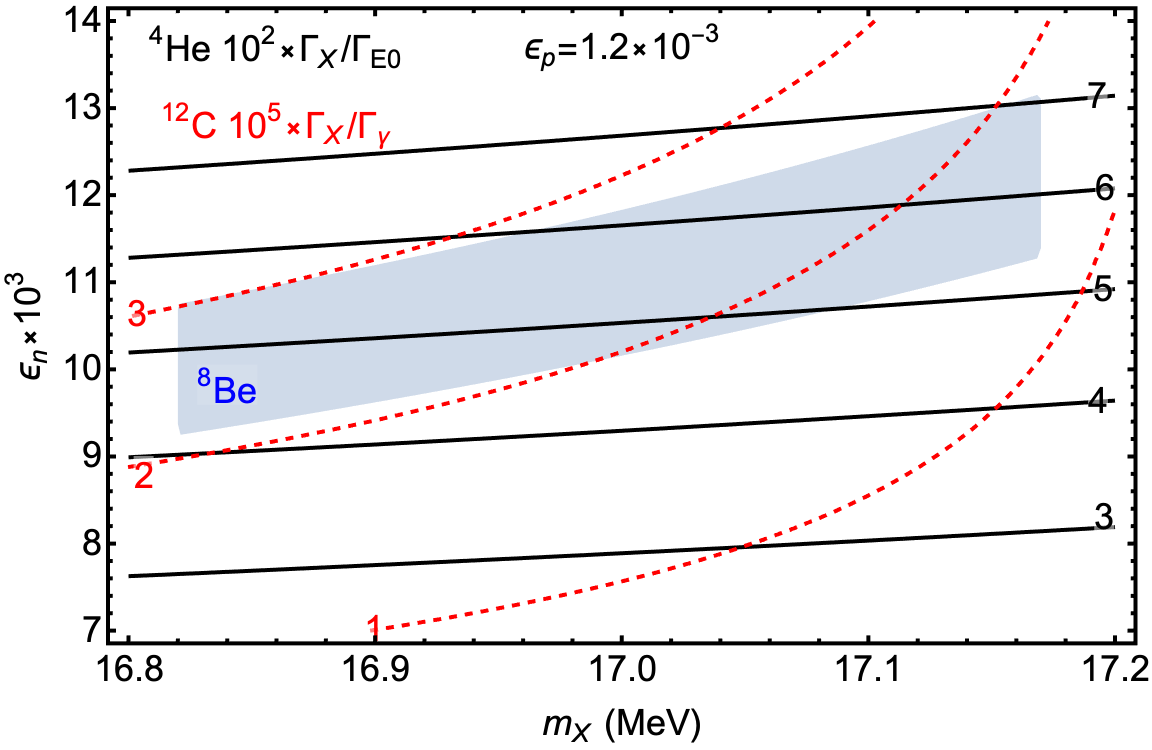} \hfil
\includegraphics[width=0.43\textwidth]{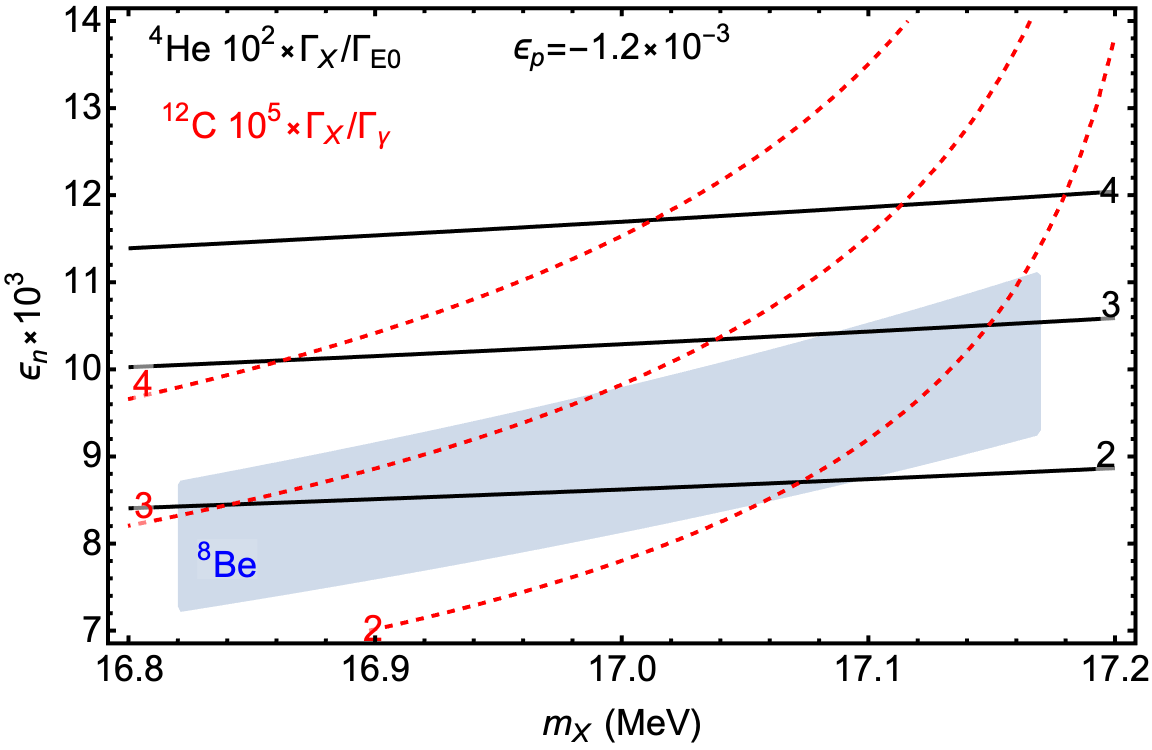} 
\vspace*{-0.1in}
\caption{Predictions for decay widths into a protophobic $X$ gauge boson in the $(\varepsilon_n, m_X)$ plane for the fixed values of $\varepsilon_p$ indicated, where $\varepsilon_n$ is the $X$ boson's coupling to neutrons, and $m_X$ is its mass.   The blue shaded regions are favored by the $^8$Be anomaly, with isospin mixing included (see \eqref{isospinmixingwidth}).   Also shown are solid black contours of $\Gamma_X/\Gamma_{\text{E0}}$ for $^4$He(20.21) decays and dashed  red contours of  $\Gamma_X/\Gamma_{\gamma}$ for $^{12}$C(17.23) decays.  
\vspace*{0.1in}
}
\label{fig:WidthPredict}
\end{figure}

\begin{figure} [tbp]
\centering
\includegraphics[width=0.43\textwidth]{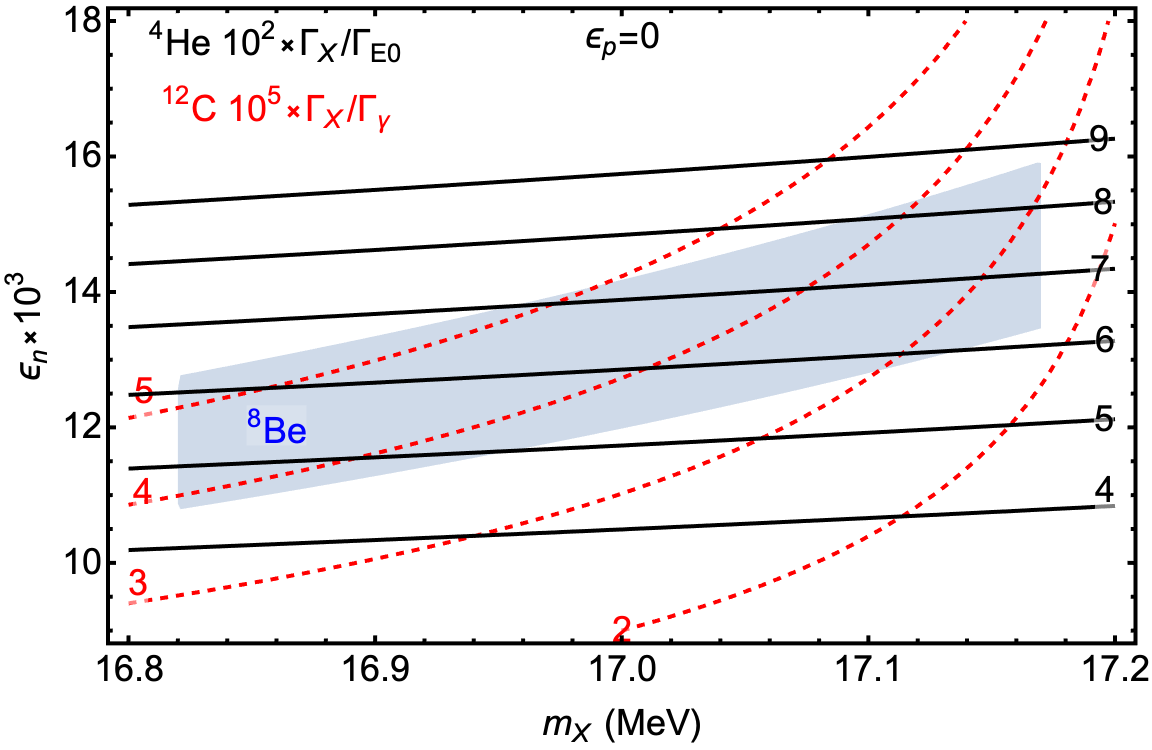} \\
\vspace*{0.1in}
\includegraphics[width=0.43\textwidth]{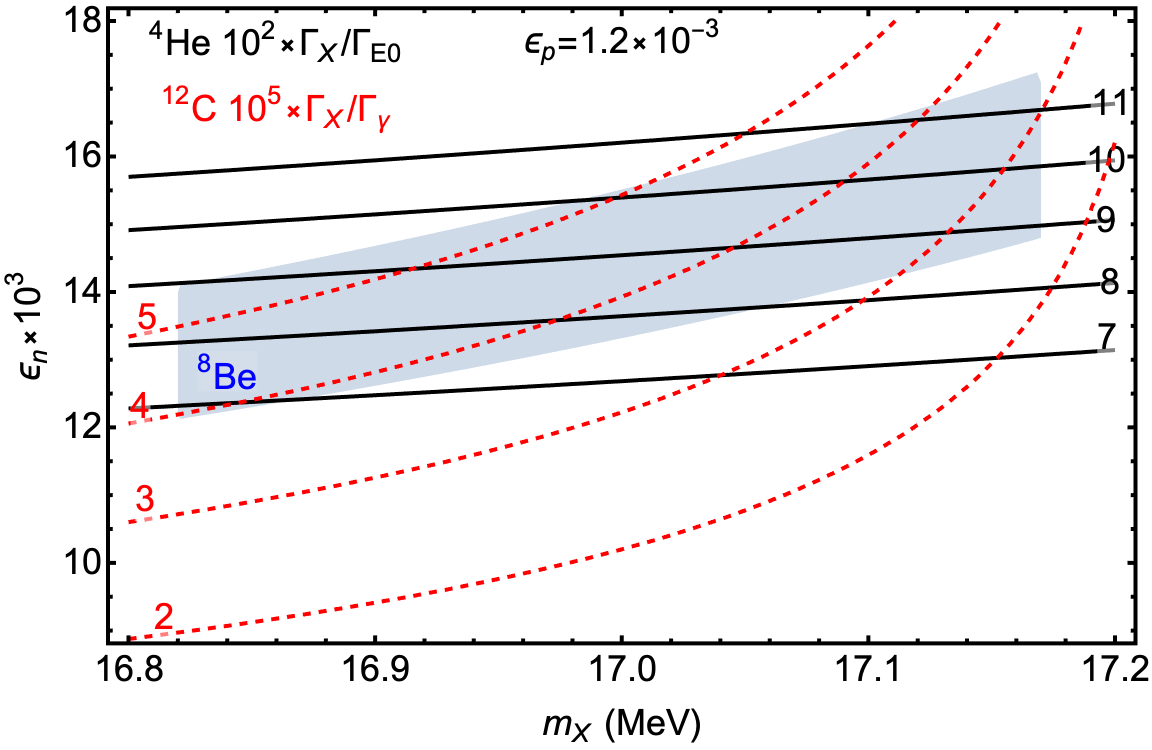} \hfil
\includegraphics[width=0.43\textwidth]{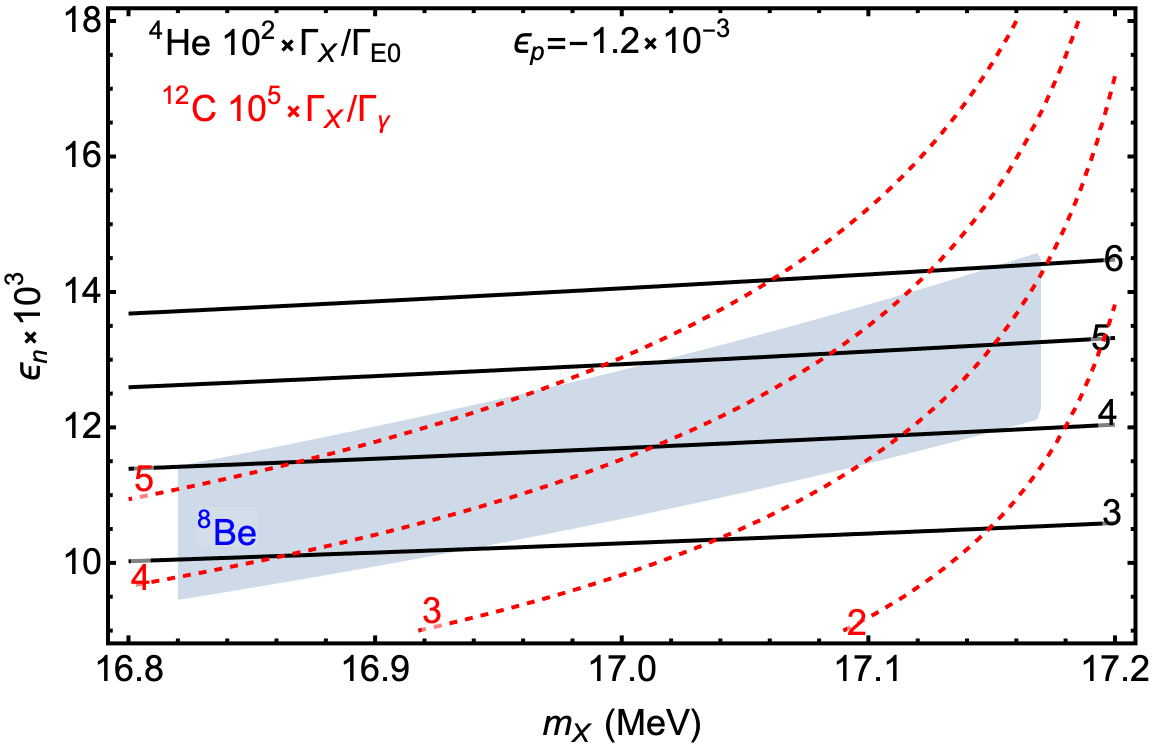} 
\vspace*{-0.1in}
\caption{As in \figref{WidthPredict}, but with both isospin mixing and isospin breaking included (see \eqref{isospinmixingbreakingwidth}). }
\label{fig:WidthPredictBreaking}
\end{figure}

In \figsref{WidthPredict}{WidthPredictBreaking}, we plot the regions of the $(m_X, \varepsilon_n)$ plane  that can explain the beryllium anomaly, with $m_X =17.01 \pm 0.16~\mev$ and~\cite{Krasznahorkay:2018snd}
\begin{equation}
\frac{\Gamma (^8\text{Be(18.15)} \to {}^8\text{Be} \, X)}{\Gamma (^8\text{Be(18.15)} \to {}^8\text{Be} \, \gamma )}
 = (6 \pm 1) \times 10^{-6} \ .
 \end{equation}
The three panels are for the purely protophobic case $\varepsilon_p = 0$ and the minimal and maximal allowed values $\varepsilon_p = \pm 1.2 \times 10^{-3}$.  In \figref{WidthPredict}, we use the theoretical prediction of \eqref{isospinmixingwidth}, which includes the isospin mixing of the beryllium excited states.  In \figref{WidthPredictBreaking}, we use the theoretical prediction of \eqref{isospinmixingbreakingwidth}, which includes both isospin mixing and isospin breaking.  The rate of the ATOMKI beryllium anomaly requires $\varepsilon_n \approx 10^{-2}$.
 
In \figsref{WidthPredict}{WidthPredictBreaking}, we also overlay black contours for $\Gamma (^4\text{He(20.21)} \to {}^4\text{He} \, X)$ normalized to the E0 decay width, using \eqref{Heliumwidth}.  We see that for vector $X$ bosons that can explain the beryllium anomaly, the predicted value of the helium decay width is 
\begin{equation}
\Gamma (^4\text{He(20.21)} \to {}^4\text{He} \, X) 
= (1-11) \times 10^{-2} \ \Gamma_{\text{E0}} 
= (0.3 - 3.6) \times 10^{-5}~\ev \ .
\label{eq:heliumprediction}
\end{equation}
This theoretical prediction overlaps with the experimentally measured range of $\Gamma (^4\text{He(20.49)} \to {}^4\text{He} \, X) = (4.0 \pm 1.2) \times 10^{-5}~\ev$ given in \eqref{width}!  The comparison at present is clouded by the experimental uncertainties regarding the E1 background discussed in \secref{introduction}, and the fact that the prediction of \eqref{heliumprediction} is for on-resonance decay, while the measurement of \eqref{width} is for off-resonance decay.  But with those uncertainties in mind, the protophobic vector boson currently provides an amazingly consistent explanation of both the beryllium and helium anomalies.  We strongly recommend that the experimental measurement be updated to include the E1 background and that a future measurement at the $^4$He(20.21) resonance be made to provide an unambiguous test of the prediction of \eqref{heliumprediction}.

Finally, in \figsref{WidthPredict}{WidthPredictBreaking}, we also overlay red contours for $\Gamma (^{12}\text{C(17.23)} \to {}^{12}\text{C} \, X)$ normalized to the photon decay width, using \eqref{Carbonwidth}.  For parameters that explain the beryllium anomaly, the carbon width is expected to be in the range
\begin{equation}
\Gamma (^{12}\text{C(17.23)} \! \to \! {}^{12}\text{C} \, X) 
= (1 - 5) \times 10^{-5} \ \Gamma (^{12}\text{C(17.23)} \! \to \! {}^{12}\text{C} \, \gamma) 
= (0.4 - 2.2) \times 10^{-3}~\ev \ .
\label{eq:carbonprediction}
\end{equation}
The expected branching ratios for decays to $X$ are similar for the beryllium and carbon cases, and an observation of these carbon decays, with the predicted branching ratio and consistent with $m_X \approx 17~\mev$, would provide overwhelming evidence for both the existing beryllium and helium anomalies and the protophobic vector boson explanation.

\section{General Spin-1 Couplings \label{sec:general}}

The ${}^4$He experimental set-up, with its ability to run at off-resonance energies, has the potential to provide additional discriminating power among possible $X$ parities. We illustrate how this works by assuming a spin-1 $X$ with {\em both} vector and axial vector couplings to quarks.  Similar remarks apply to the case of a spin-0 $X$ boson with both scalar and pseudoscalar interactions, although, as noted at the end of \secref{pseudoscalar}, spin-0 states with mixed parity are now strongly disfavored as explanations of the ${}^8$Be and ${}^4$He anomalies.

If the $X$ boson is a mixture of vector and axial vector, it could generically be produced by both the $0^-$ and $0^+$ $^4$He excited states.  For the purposes of this analysis, we consider the true excited state $N_\ast$ to be a linear combination of $0^-$ and $0^+$.  The full production cross section is then split up into $\sigma_-$ and $\sigma_+$ with
\beq
\sigma_\pm\equiv \sigma(p+{}^3\text{H}\to 0^\pm) \ .
\eeq

Since $X$ is assumed to be produced through both states, the complete $X$ production cross section is
\beq
\sigma_X=\sigma_{-}\frac{\Gamma\left(0^-\to X \right)}{\Gamma_-}+\sigma_{+}\frac{\Gamma\left(0^+\to X \right)}{\Gamma_+} \ ,
\eeq
where $\Gamma_\pm$ is the total width of the $0^\pm$ excited state. Because the experimental signal is normalized to the E0 electromagnetic transition, which occurs only through the $0^+$ state, we define $\sigma_\text{E0}$ in a similar way as 
\beq
\sigma_\text{E0}=\sigma_+\frac{\Gamma_\text{E0}}{\Gamma_+} \ .
\eeq
Consequently, the ratio of the production cross sections is
\beq
\frac{\sigma_X}{\sigma_\text{E0}}=\frac{\Gamma\left(0^+\to X \right)}{\Gamma_\text{E0}}+\frac{\sigma_- \, \Gamma_+}{\sigma_+ \, \Gamma_-} \,\frac{\Gamma\left(0^-\to X \right)}{\Gamma_\text{E0}} \ .
\eeq
This ratio is proportional to the number of $X$ decay events divided by the number of E0 events recorded by the experiment.  The goal is to extract the $X$ decay widths of the two individual states, with the E0 width and the two total widths $\Gamma_\pm$ known experimentally. The more subtle quantity is $\sigma_-/\sigma_+$.

One might hope to obtain these cross sections from the widths by using the relation
\beq
\sigma\left(p+{}^3\text{H}\to N_\ast\right)=\frac{4 \pi^2 (2J_\ast+1)}{M_\ast} \, \Gamma\left(N_\ast\to p+{}^3\text{H} \right) \delta(E_\text{CM}^2-M_\ast^2) \ ,
\eeq
treating $N_\ast$ as a bound state of $p$ and ${}^3$H with mass $M_\ast$ and spin $J_\ast$.  However, for off-shell running, one should broaden the $\delta$-function into the resonance peak. This can be approximated by taking the narrow width approximation in reverse,
\beq
\delta(E_\text{CM}^2-M_\ast^2)\to\frac{M_\ast \Gamma_\ast/\pi}{(E_\text{CM}^2-M_\ast^2)^2+M_\ast^2\Gamma_\ast^2} \ ,
\eeq
leading to the relation
\beq
\sigma\left(p+{}^3\text{H}\to N_\ast\right)=\frac{4 \pi (2J_\ast+1) \Gamma_\ast}{(E_\text{CM}^2-M_\ast^2)^2+M_\ast^2\Gamma_\ast^2} \, \Gamma\left(N_\ast\to p+{}^3\text{H} \right) .
\eeq
In this result, $\Gamma_\ast$ is the full width of the bound state, whereas the particular production mode may only be related to a particular partial width. The $0^+$ nearly always decays to protons, but the $0^-$ decays to proton final states only 76\% of the time~\cite{Wang:2017}.  We therefore obtain
\beq
\frac{\sigma_-}{\sigma_+}=\frac{0.76 \, \Gamma_-^2 }{\Gamma_+^2}\frac{\left(E_\text{CM}^2-M_{+}^2\right)^2+M_{+}^2\Gamma_{+}^2}{\left(E_\text{CM}^2-M_{-}^2\right)^2+M_{-}^2\Gamma_{-}^2} \ ,
\eeq
where $M_{\pm}$ is the nuclear mass of the $0^\pm$ excited state.  Putting everything together, we find
\beq
\frac{\sigma_X}{\sigma_\text{E0}}=\frac{\Gamma\left(0^+\to X \right)}{\Gamma_\text{E0}}+\frac{0.76 \, \Gamma_- }{\Gamma_+} \, \frac{\left(E_\text{CM}^2-M_{+}^2\right)^2+M_{+}^2\Gamma_{+}^2}{\left(E_\text{CM}^2-M_{-}^2\right)^2+M_{-}^2\Gamma_{-}^2} \, \frac{\Gamma\left(0^-\to X \right)}{\Gamma_\text{E0}} \ ,
\eeq
which relates the total number of $X$ events observed at a given $E_\text{CM}$ to the individual partial widths $\Gamma\left(0^+\to X \right)$ and $\Gamma\left(0^-\to X \right)$.  

As can be seen in \figref{BreitWig}, by varying the proton beam energy $E_{\text{beam}}$, one can scan over the $0^+$ peak.  By measuring the variation of $\sigma_X/\sigma_\text{E0}$ in such a scan, which requires the proper simulation of all relevant backgrounds, the two widths $\Gamma\left(0^\pm\to X \right)$ can be extracted.  In particular, if the ratio remains constant, it would suggest that $\Gamma\left(0^-\to X \right)=0$, implying that the $X$ boson is a pure vector.  Such a result would also exclude the possibility of $X$ being a pseudoscalar.

\section{Conclusions \label{sec:conclusions}}

The transitions between energy levels of light nuclei are a natural laboratory to explore $\sim$~MeV mass particles with ultra-weak interactions with the standard model.  These nuclear transitions are able to probe models that are interesting from the point of view of theories of dark matter and dark forces, and they provide information complementary to searches for such particles from accelerator and astrophysical sources. 

Several years ago, the ATOMKI group discovered a resonance structure in the invariant mass of $e^+ e^-$ pairs produced in the transition of an excited state of ${}^8\text{Be}$ to its ground state.  The results have so far defied plausible explanation through prosaic nuclear physics and are suggestive of a new particle with a mass around 17~MeV.  Since then, the new particle interpretation has been bolstered by the discovery of viable new particle explanations that are consistent with all other experimental constraints.  Most stunningly, the ATOMKI group has recently confirmed their original results, both in ${}^8\text{Be}$, using new detector elements, and by new observations in a transition of ${}^4\text{He}$ whose kinematics point to the same 17 MeV mass required to explain the ${}^8\text{Be}$ results.

In this work, we provide a comprehensive theoretical analysis of these anomalies by examining the dynamical consistency of the size of the two observed excesses.  We construct an effective field theory to describe the nuclear transitions of interest, and we examine the possibility that the new particle $X$ is a scalar, pseudoscalar, vector, or axial vector boson coupling to nucleons and electrons.  We find that it is very difficult to simultaneously explain the ${}^8$Be and ${}^4$He results with a spin-0 particle of either parity (or a mixture).  For the axial vector case, the EFT predicts that the ${}^8$Be and ${}^4$He decay widths differ by a factor of $\sim 10^2$, in contrast to the observations, which find that they are similar, but this discrepancy could conceivably be reconciled by significant uncertainty in nuclear matrix elements.  

In the vector case, however, precise predictions can be made that are independent of nuclear physics matrix elements by normalizing them to known EM transitions.  The result is that if the $X$ boson is a vector particle, the protophobic couplings required by the ${}^8\text{Be}$ results predict rates for the ${}^4\text{He}$ transitions with no free parameters.  The results are
\begin{eqnarray}
\text{Protophobic vector boson: } && \Gamma (^4\text{He(20.21)} \to {}^4\text{He} \, X) 
= (0.3 - 3.6) \times 10^{-5}~\ev \qquad \\
\text{ATOMKI Experiment~\cite{Krasznahorkay:2019lyl,Krasznahorkay_He_Proceedings}: } && \Gamma (^4\text{He(20.21)} \to {}^4\text{He} \, X) 
= (2.8 - 5.2) \times 10^{-5}~\ev .
\end{eqnarray}
{\em The reported 7$\sigma$ anomalies reported in $^8$Be and $^4$He nuclear decays are both kinematically and dynamically consistent with the production of a 17 MeV protophobic gauge boson.}

What is the path forward? Clearly, now is the time for other collaborations to perform the same nuclear measurements to check the ATOMKI results. But in this work, we also propose simple modifications of the ATOMKI setup that could provide incisive tests of the new particle interpretation.  The comparison between theory and experiment will be sharpened considerably by including the E1 background in the experimental analysis and running on the $^4$He(20.21) $0^+$ resonance.  In addition, scanning through the $^4$He(20.21) $0^+$ resonance can provide important information to disentangle vector and axial vector $X$ bosons and quantify the properties of particles with mixed couplings. Last, we find that the protophobic vector boson could also be observable in the decays of the $^{12}\text{C}$(17.23) $1^-$ excited state, and we have provided precise predictions for this rate. 

The results of this study therefore lay out a roadmap for possible future measurements that will further shed light on the ATOMKI anomalies.  If the predictions are confirmed, these measurements will provide overwhelming evidence that a fifth force has been discovered.

\acknowledgments

We thank Bart Fornal, Iftah Galon, Susan Gardner, and Attila Krasznahorkay for helpful correspondence. This work is supported in part by U.S.~National Science Foundation Grant No.~PHY-1915005.  The work of JLF and CBV is supported in part by Simons Investigator Award \#376204. 

\appendix

\bibliography{fifthforcebib}

\end{document}